\documentclass[%
 reprint,
 superscriptaddress,
 amsmath,amssymb,
 aps,
]{revtex4-2}

\usepackage{graphicx}% Include figure files
\usepackage{bm}% bold math
\usepackage{xurl}
\newcommand{\Npop}{N^{\mathrm{(pop)}}}

\begin{document}

\title{COVID-19 lockdown induces disease-mitigating structural changes in mobility networks}% Force line breaks with \\

\author{Frank Schlosser}
\affiliation{Robert Koch-Institute, Nordufer 20, D-13353 Berlin, Germany}
\affiliation{Institute for Theoretical Biology, Humboldt-University of Berlin, Philippstr.\,13, D-10115 Berlin, Germany}
\author{Benjamin F. Maier}
\affiliation{Robert Koch-Institute, Nordufer 20, D-13353 Berlin, Germany}
\author{Olivia Jack}
\affiliation{Robert Koch-Institute, Nordufer 20, D-13353 Berlin, Germany}
\author{David Hinrichs}
\affiliation{Robert Koch-Institute, Nordufer 20, D-13353 Berlin, Germany}
\author{Adrian Zachariae}
\affiliation{Robert Koch-Institute, Nordufer 20, D-13353 Berlin, Germany}
\author{Dirk Brockmann}
\affiliation{Robert Koch-Institute, Nordufer 20, D-13353 Berlin, Germany}
\affiliation{Institute for Theoretical Biology, Humboldt-University of Berlin, Philippstr.\,13, D-10115 Berlin, Germany}

% \affiliation{
% Robert Koch-Institute, Nordufer 20, D-13353 Berlin, Germany
%}
%\affiliation{Institute for Theoretical Biology, Humboldt-University of Berlin, Philippstr. 13, D-10115 Berlin, Germany}

% \author{Frank Schlosser$^{1,2}$, Benjamin F. Maier$^1$, David Hinrichs$^1$, Adrian Zachariae$^1$, Dirk Brockmann$^{1,2}$}

% \affiliation{
%     $^1$Robert Koch-Institute, Nordufer 20, D-13353 Berlin, Germany
% }
% \affiliation{
%     $^2$Institute for Theoretical Biology, Humboldt-University of Berlin, Philippstr. 13, D-10115 Berlin, Germany
% }

%\author[a, b]{Frank Schlosser}
%\author[a]{Benjamin F. Maier} 
%\author[a]{David Hinrichs}
%\author[a]{Adrian Zachariae}
%\author[a, b]{Dirk Brockmann}

%\affil[a]{Robert Koch-Institute, Nordufer 20, D-13353 Berlin, Germany}
%\affil[b]{Institute for Theoretical Biology, Humboldt-University of Berlin, Philippstr. 13, D-10115 Berlin}

\date{\today}

\begin{abstract}
In the wake of the COVID-19 pandemic many countries implemented containment measures to reduce disease transmission. Studies using digital data sources show that the mobility of individuals was effectively reduced in multiple countries. However, it remains unclear whether these reductions caused deeper structural changes in mobility networks, and how such changes may affect dynamic processes on the network. Here we use movement data of mobile phone users to show that mobility in Germany has not only been reduced considerably: Lockdown measures caused substantial and long-lasting structural changes in the mobility network. We find that long-distance travel was reduced disproportionately strongly. The trimming of long-range network connectivity leads to a more local, clustered network and a moderation of the ``small-world'' effect. We demonstrate that these structural changes have a considerable effect on epidemic spreading processes by ``flattening'' the epidemic curve and delaying the spread to geographically distant regions.
\end{abstract}

% At least three keywords are required at submission. Please provide three to five keywords, separated by the pipe symbol.

\maketitle

% \dates{This manuscript was compiled on \today}
% \doi{\url{www.pnas.org/cgi/doi/10.1073/pnas.XXXXXXXXXX}}

% \thispagestyle{firststyle}
% \ifthenelse{\boolean{shortarticle}}{\ifthenelse{\boolean{singlecolumn}}{\abscontentformatted}{\abscontent}}{}

% \todo{I recommend removing Germany from the title.}

% Introduction
\section{Introduction}

During the first phase of the coronavirus disease 2019 (COVID-19) pandemic, countries around the world implemented a host of containment policies aimed at mitigating the spread of the disease \cite{Wilder-Smith2020, Sohrabi2020, Zhang, Fisher}. Many policies restricted human mobility, intending to reduce close-proximity contacts, the major driver of the disease's spread \cite{Li2020}. In Germany, these  policies included border closures and travel bans, restrictions of public activity (school and business closures), paired with appeals by the government to avoid trips voluntarily whenever possible \cite{ACAPS}. We will refer to these policies as ``lockdown'' measures for brevity.

Based on various digital data sources such as mobile phone data or social media data, several studies show that mobility significantly changed during lockdowns \cite{Oliver2020}. Most studies focused on general mobility trends and confirmed an overall reduction in mobility in various countries\cite{Klein2020, Lee2020, Pepe2020, Gao2020, Pullano2020}. Other research focused on the relation between mobility and disease transmission: For instance, it has been argued that mobility reduction is likely instrumental in reducing the effective reproduction number in many countries \cite{Flaxman2020, Dehning2020, Yabe2020, Lemaitre2020, Jia2020}, in agreement with theoretical models and simulations, which have shown that containment can effectively slow down disease transmission \cite{Maier2020, Arenas2020, Chinazzi2020}.

However, it remains an open question whether the mobility restrictions promoted deeper structural changes in mobility networks, and how these changes impact epidemic spreading mediated by these networks. Recently,  Galeazzi~\emph{et al.}~ \cite{Galeazzi2020} found increased geographical fragmentation of the mobility network. A thorough understanding of how structural mobility network changes impact epidemic spreading is needed in order to correctly assess the consequences of mobility restrictions not only for the current COVID-19 pandemic, but also for for similar scenarios in the future.

Here, we analyze structural changes in mobility patterns in Germany during the COVID-19 pandemic. We analyze movements recorded from mobile phones of $43.6$~million individuals in Germany. Beyond a general reduction in mobility, we find considerable structural changes in the mobility network. Due to the reduction of long-distance travel, the network becomes more local and lattice-like. Most importantly, we find a changed scaling relation between path lengths and geographic distance: During lockdown, the effective distance (and arrival time in spreading processes) to a destination continually grows with geographic distance. This shows a marked reduction of the ``small-world'' characteristic, where geographic distance is usually of lesser importance in determining path lengths \cite{Watts1998, Brockmann2013}. Using simulations of a commuter-based SIR model, we demonstrate that these changes have considerable practical implications as they suppress (or ``flatten'') the curve of an epidemic remarkably and delay the disease's arrival between distant regions.

\section{Mobility trends in Germany}
\label{sec:mob_reduction}

\begin{figure*}[ht]
\centering
\includegraphics[width=1.0\linewidth]{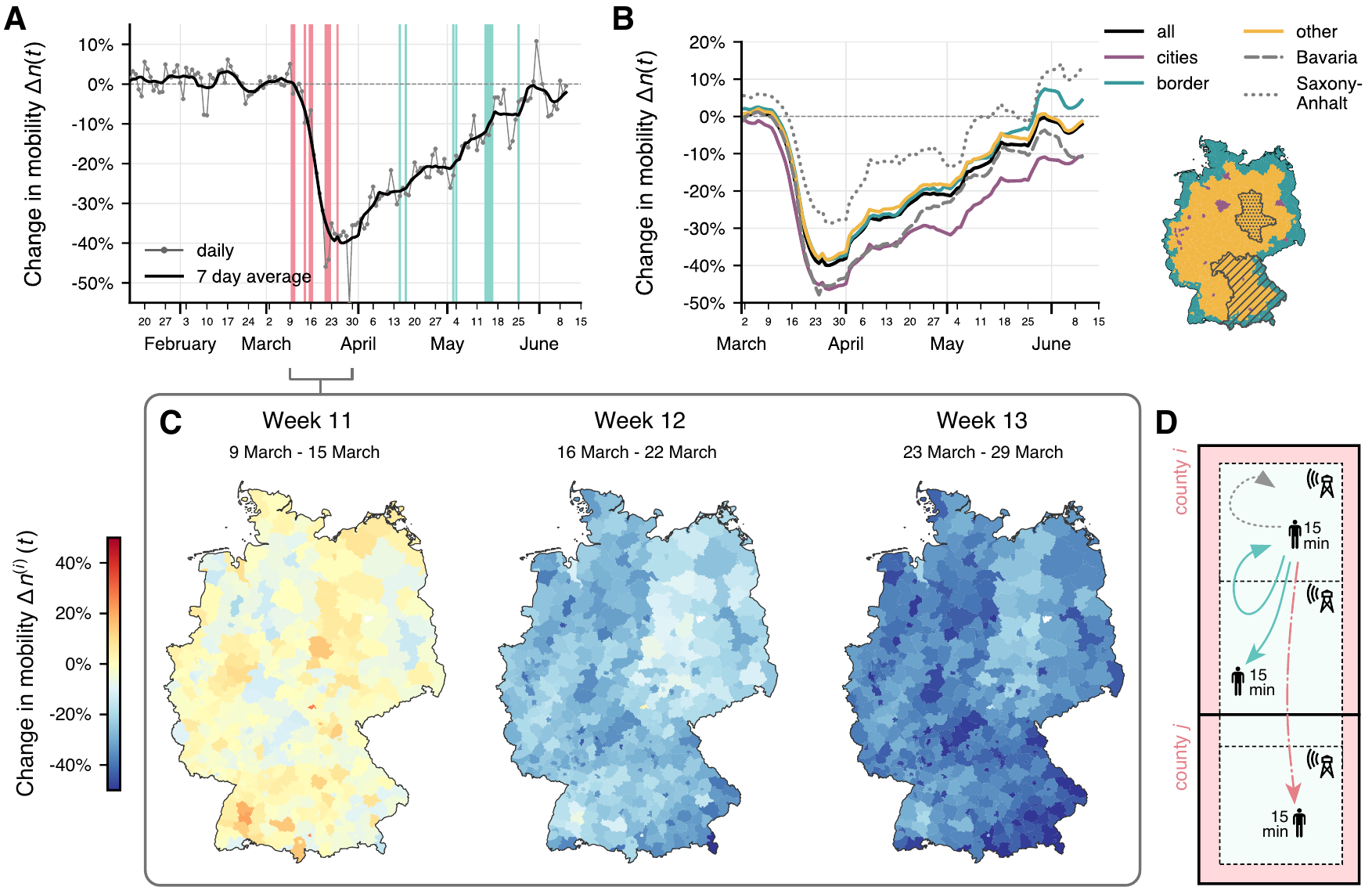}
\caption{\label{fig:mobility_reduction} Mobility changes in Germany during the COVID-19 pandemic. (A) The change in total movements $\Delta n(t)$ in 2020, relative to March 2019. Mobility decreases drastically in mid-March, coincident with restricting measure implementations (red bars), followed by a gradual increase in mobility concurrent with the lifting of restricting policies (teal bars). Bar width indicates the number of policies issued or lifted on that date, respectively. (B) The change in mobility is spatially heterogeneous. Mobility is reduced more in large cities (shown here for the 20 largest cities) and states that implemented more severe restrictions (such as Bavaria). (C) Mobility change in German counties for the three weeks with most substantial global change. The mobility change $\Delta n^{(i)}(t)$ represents the number of trips that originate in county $i$ (see \emph{Materials and Methods}). (D) Illustration of how mobility is recorded. A trip is counted when a user switches to one or multiple new cell towers, until the user becomes stationary again (no further switch for approx. 15 minutes). Trips can be within the same county (teal, solid line) or between different counties (red, dot-dashed line). Movements without changing cell towers are not recorded (grey, dotted line).}
\end{figure*}

\subsection{General mobility changes}
\label{sec:general_mobility}

We base our analysis on mobility flows collected from mobile phone data. The data counts the number of \emph{trips}, where a trip is defined as a single mobile phone switching cell towers at least once, between two resting phases of at least 15 minutes (see Fig.~\ref{fig:mobility_reduction}D). A resting phase is defined as a mobile phone not switching its connected cell tower. These trips are aggregated over the course of a day to build the daily flow matrix $\mathcal{F}(t)$. The element $F_{ji}(t)$ quantifies the total number of trips from location $i$ to location $j$ on a given day $t$. Locations are the $m=401$ counties of Germany. Note that flows within counties $F_{ii}(t)$ are included. During times with normal mobility, (e.g.~during March 2019, which we use as a baseline, see below) the total flow is $176~\mathrm{million}$ trips per day on average, recorded among $43.6$ million users \cite{Telefonica}, corresponding to an average of $3.8$ trips per user per day. The baseline average daily flow between all pairs of locations is $\langle F_{ji}(t)\rangle=1103$ (averaged over all days in March 2019) with a standard deviation of $\mathrm{Std}[F_{ji}(t)]=26.413$. Flows below a threshold of $F_{ji}(t)<5$ were omitted from the data due to anonymization requirements. 

To analyze general changes in mobility during the COVID-19 pandemic, we focus on the \emph{daily mobility change} $\Delta n(t)$, which is the relative difference in the total number of trips $N(t)=\sum_{i,j=1}^m F_{ji}(t)$ compared to the baseline number of trips, i.e. during a period of ``normal'' mobility. Here, we use March 2019 as this baseline period, and compare the mobility on each date $t$ from 2020 to the average mobility on the corresponding weekday in March 2019 (see \emph{Materials and Methods}).

We find that mobility in Germany was substantially reduced during the COVID-19 pandemic, see Fig.~\ref{fig:mobility_reduction}. The largest reduction occurred in mid-March, when the vast majority of mobility-reducing interventions took effect (information on government policies is taken from the ACAPS dataset, see SI). Over the course of three weeks, mobility dropped to $-40\%$ below baseline  on March 27th in the 7 day moving average. The total number of daily recorded trips decreased from $176\,\mathrm{million}$ to $107\,\mathrm{million}$ trips (from $3.8$ to $2.3$ daily trips per user). The decline was followed by an immediate rebound at the beginning of April, even though mobility-restricting regulations remained in effect during this period. In the following months, mobility increased slowly, reaching its pre-lockdown levels in early June. Interestingly, the increase in mobility took place in small bursts followed by short periods of stagnation. These bursts started at around the same time that mobility-restricting policies were lifted, hinting at a causal relationship.

Mobility did not decrease homogeneously in Germany: Some areas witnessed a more substantial reduction than others. We observed a greater mobility reduction in Western and Southern states (such as Bavaria), which were more substantially affected by the pandemic, compared to the Eastern states of Germany (for example in Saxony-Anhalt) \cite{RobertKoch-Institute2020}. This difference can partially be explained by more severe mobility restrictions in some Western states. For instance, Bavaria passed stricter measures on May 20th, resulting in a higher reduction in mobility in calendar week 13. Still, most policies were uniform across Germany and were implemented in a similar manner on a federal level. Therefore, differences in policies can only deliver a partial explanation for regional heterogeneities.
%We found that the spatially heterogeneous impact of COVID-19 can partially account for the observed differences: Regions with more cases of COVID-19 show a greater reduction in mobility (see supplement \ref{app:cases_mobility_correlation}).
Furthermore, we found systematic dependencies on demographic factors. Mobility is reduced more in large cities compared to less densely populated areas. In addition, several border regions particularly associated with cross-border traffic exhibit a higher than average mobility reduction , although the border as a whole does not deviate markedly from the average.

\subsection{Distance-dependence of mobility reduction}

\begin{figure}[ht]
\centering
\includegraphics[width=1.0\linewidth]{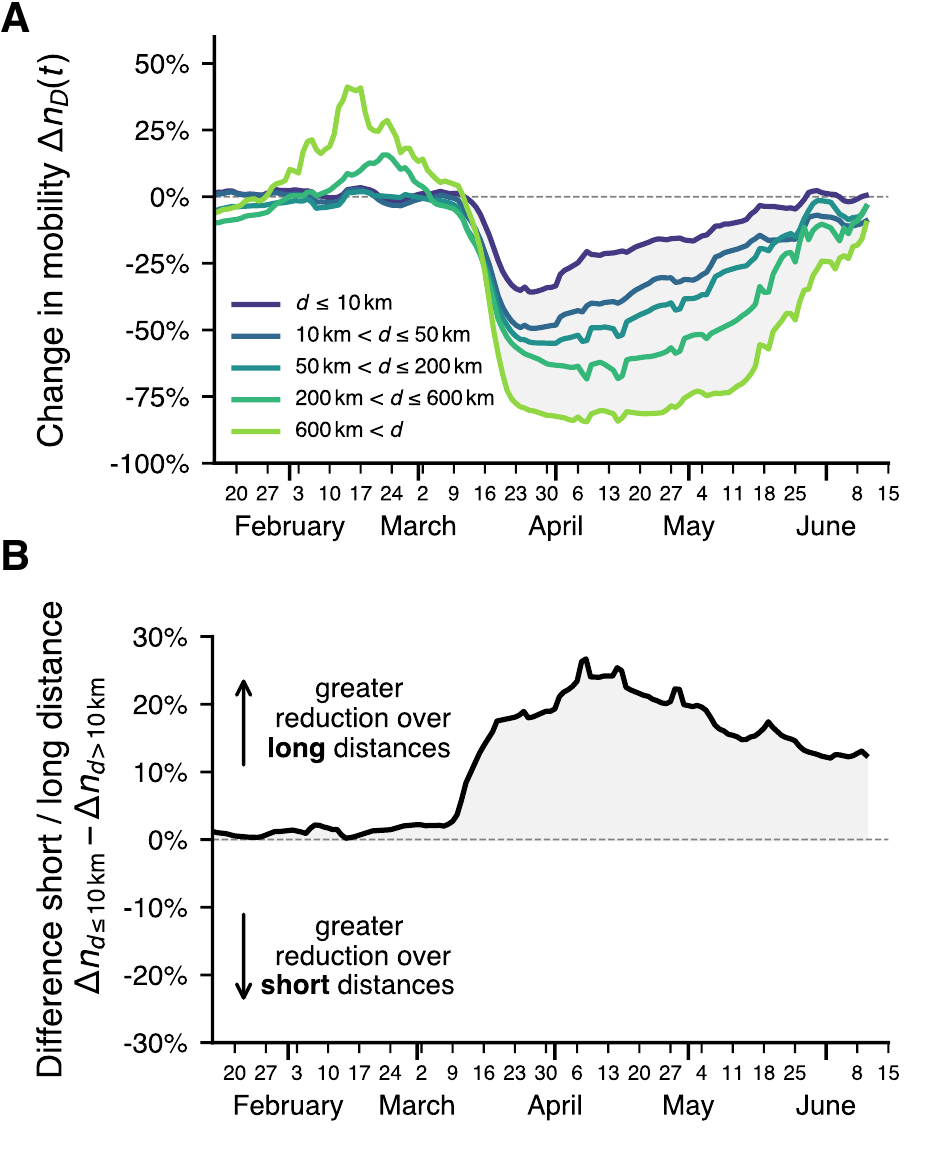}
\caption{\label{fig:distance_reduction} Mobility reduction as a function of distance. (A) Relative mobility changes $\Delta n_D(t)$ for different distance ranges $D$ (7 day moving averages). Long-distance trips reduction is higher. The notable increase in long-distance trips in February coincides with school holidays in several German states. Fluctuations in April and May are often centered around public holidays. (B) The difference between short-distance mobility change $\Delta n_{d\leq10\,\mathrm{km}}(t)$ and long-distance change $\Delta n_{d>10\,\mathrm{km}}(t)$ is a useful indicator for unusual mobility.}
\end{figure}

The observed general reduction in mobility begs the question of \emph{how} mobility has changed, and what types of trips were reduced. We observe a distinct dependence of mobility change on trip length, see Fig.~\ref{fig:distance_reduction}. We calculated the mobility change $\Delta n_D(t)$ for all trips in a certain distance range $D$. Because data is aggregated on a county level, we use the distance between the county centroids as an estimate of trip distance (see \emph{Materials and Methods}).

Over the full range of observed distances, we find that long-distance trips decreased more strongly than short-distance trips. This resonates with the expectation that many social-distancing policies targeted long-distance travel specifically: Travel bans across country and state borders, cancellations of major events, and border closures by other countries affecting holiday travel.

Furthermore, we find that the split between short- and long-distance mobility reduction is a useful indicator for an unusual state of the mobility network. While the total number of trips has almost returned to its pre-pandemic state (see Fig.~\ref{fig:mobility_reduction}A), which could at first glance give the impression that normal mobility patterns have been restored, the continued discrepancy between short- and long-distance mobility reduction indicates a long-lasting structural change in mobility patterns (see Fig.~\ref{fig:distance_reduction}). The discrepancy, while declining slightly, remained stable over the course of the pandemic, evidence for the prevalent impact of mobility changes.

\section{Structural changes in the mobility network}

\subsection{Pre-lockdown and lockdown mobility networks}

\begin{figure*}
\centering
\includegraphics[width=1.0\linewidth]{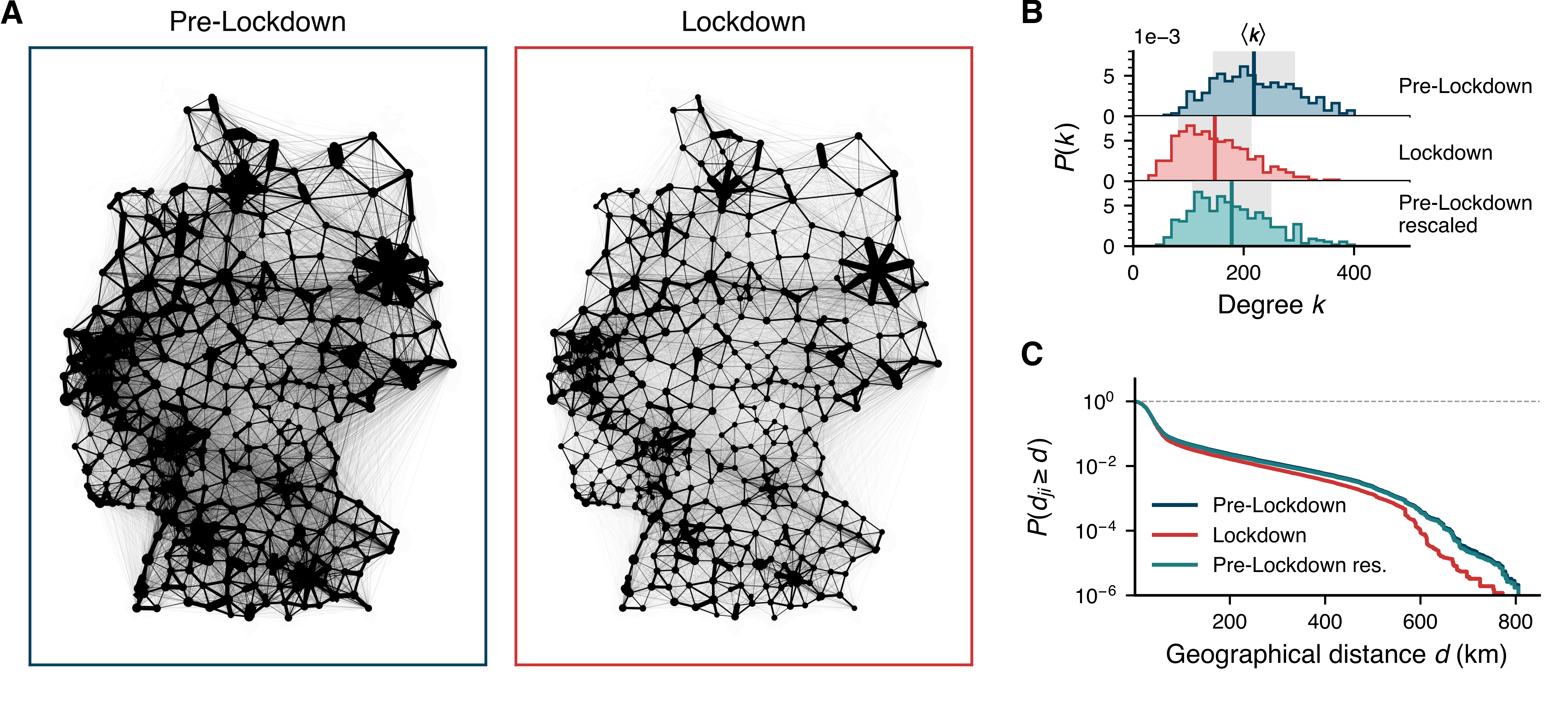}
\caption{\label{fig:networks} Comparison of the pre-lockdown mobility network $G_{10}$ and the lockdown network $G_{13}$ corresponding to calendar weeks 10 and 13, respectively. (A) Depiction of the networks. Line widths indicate the average number of daily trips along each connection. During lockdown, there are less trips in total, less unique edges, and fewer long-distance connections. (B) Distribution of node degrees $k$. The average node degree $\langle k \rangle$ and network density $\rho$ are lower during lockdown ($\langle k \rangle=148, \rho=32.6\%$) than in the pre-lockdown network ($\langle k \rangle=219, \rho=49.8\%$). These differences are only partially explained by a uniform, global reduction of trips (which causes trips to fall below the anonymization threshold $w^c=5$), as demonstrated by comparison to the rescaled pre-lockdown network $G_{10}^*(T=13)$ which is structurally similar to network $G_{10}$ but has the same number of total trips as $G_{13}$ (see \emph{Materials and Methods}). (C)  The probability $P(d_{ji}\geq d)$ that a randomly chosen edge $w_{ji}$ is of a distance $d_{ji}\geq d$. The lockdown network contains considerably fewer long-distance trips than the pre-lockdown network, an effect that cannot be explained by a uniform, global reduction of flows (rescaled pre-lockdown network).}
\end{figure*}

To identify key structural changes over time, we analyze the \emph{mobility networks} $G_T$ for each calendar week $T$, where the edge weights $w_{ji}$ correspond to the average daily flow along the edge during this week (see \emph{Materials and Methods}). To highlight the changes occurring during lockdown, we compare two specific time periods: The \emph{pre-lockdown} network $G_{10}$ is constructed from the trips in calendar week $10$ (March 2-8), before policy interventions were passed. The \emph{lockdown} network $G_{13}$, is constructed from all trips in calendar week 13 (March 23-29, the week with the highest reduction in mobility).  

The lockdown network $G_{13}$ is considerably less dense than the pre-lockdown network $G_{10}$, see Fig.~\ref{fig:networks}. Many pairs of counties with traffic under normal conditions lack traffic during the lockdown week or, the average daily flow fell below the anonymization threshold $w^c=5$. In particular, the lockdown network has fewer long-distance flows than the pre-lockdown network (Fig.~\ref{fig:networks}C), in line with our previous finding that mobility over long distances was reduced most substantially (compare Fig.~\ref{fig:distance_reduction}).

The loss of density during lockdown cannot be explained by a global, uniform reduction of mobility alone, which causes trips to fall below the observation threshold $w^c=5$. To illustrate this point, we compare the lockdown network $G_{13}$ to the rescaled network $G_{10}^*(T=13)$ where edge weights of the pre-lockdown network $G_{10}$ were rescaled such that it is structurally similar to the pre-lockdown network $G_{10}$ but has the same total number of trips as the lockdown network $G_{13}$ (see \emph{Materials and Methods}). This rescaling can rule out effects that originate in a homogeneous, global mobility reduction. We find that the rescaled pre-lockdown network is denser than the lockdown network. Specifically, we find a greater probability of observing long-distance travel (see (Fig.~\ref{fig:networks}, panels B and C). We conclude that long-distance travel has been reduced more substantially during lockdown than can be explained by a mere global reduction of mobility and thresholding effects.

\subsection{Lockdown effects on path lengths in the networks}

\begin{figure*}[ht]
\centering
\includegraphics[width=1.0\linewidth]{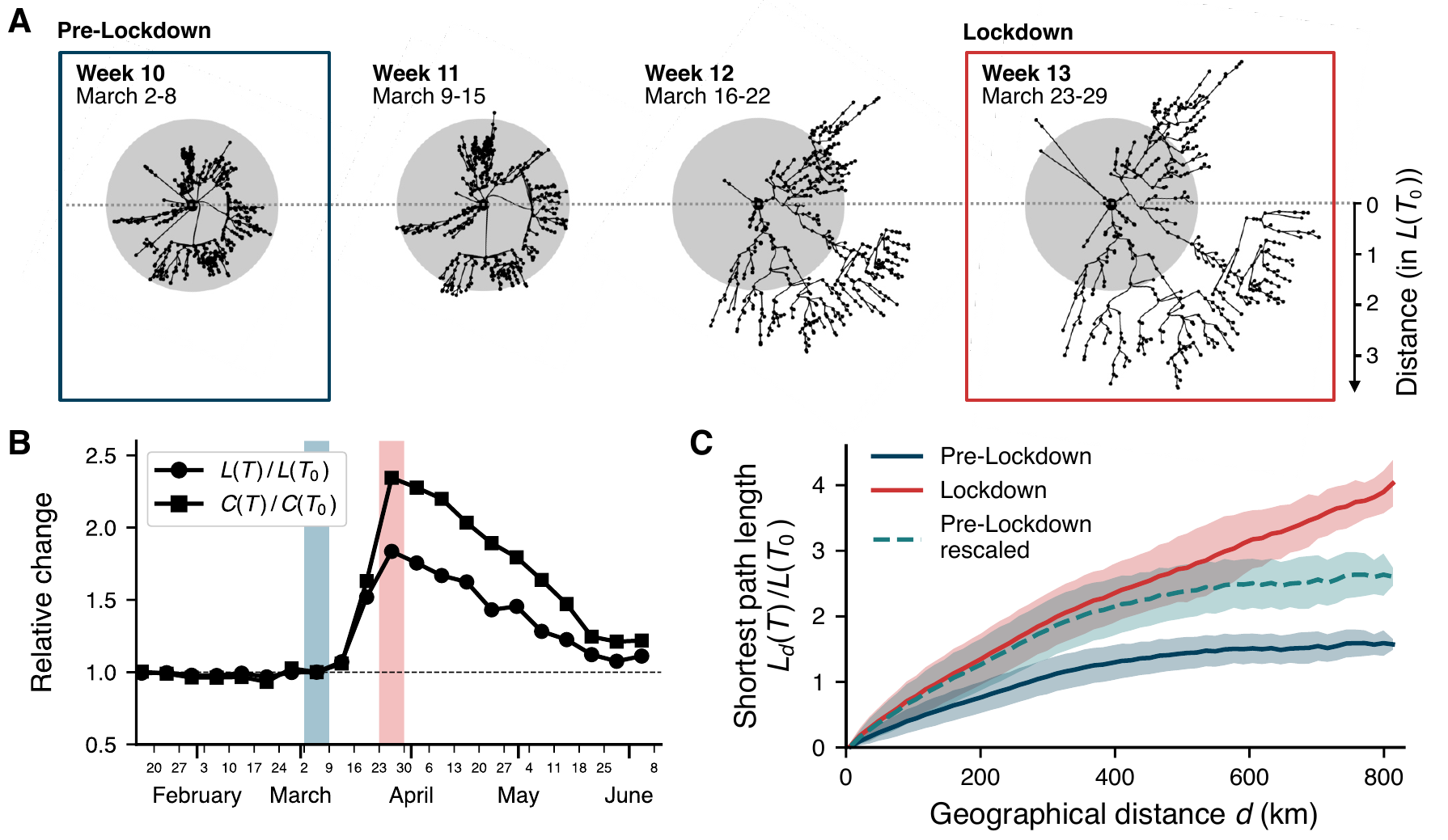}
\caption{\label{fig:networks_measures} Lockdown effects on structural network metrics. (A) The shortest path tree originating at Berlin for the weekly mobility networks $G_T$. In the pre-lockdown network $G_{10}$ (week $T_0=10$, blue frame), long-distance connections facilitate quick traversals. In the lockdown network $G_{13}$ (week $T=13$, red frame), shortest paths are generally longer and include more local steps between neighboring counties. Radial distance is scaled in multiples of the average shortest path length $L(T_{0})$ in week $T_0=10$. Grey circles mark the largest shortest path length in week 10. Further plots for Berlin and for other sources (which we find to exhibit qualitatively similar changes) are provided in the SI, section 3D. (B) The average shortest path length $L(T)$ and the average clustering coefficient $C(T)$ for weekly mobility networks $G_T$ over time, relative to their values in week $T_0=10$ (blue bar). Both metrics increase substantially in the following weeks and peak for the lockdown network $G_{13}$ (red bar), indicating a more clustered and sparser network. (C) The expected shortest path length $L_d(T)$ at distance $d$, i.e.~$L_d(T)=\langle L_{ji}(T) | d_{ji}\in\left[ d-\epsilon, d+\epsilon \right]\rangle$.  In the pre-lockdown network $G_{10}$, the shortest path length $L_{d}$ is independent of geographical distances $d$ at large distances, a known phenomenon of spatial small-world networks. In contrast, we observe a continued, roughly linear, scaling relation for $L_{d}$ in that distance range for the lockdown network $G_{13}$, a known property of lattices. The rescaled, pre-lockdown network $G_{10}^*(T=13)$ does not replicate the changed scaling behavior, demonstrating that the effect is not solely explained by a global, uniform reduction of mobility and thresholding effects.}
\end{figure*}
%\todo{it's unclear how the clustering coefficient is defined in the text of Fig 4.} 

The structural mobility changes during lockdown impact properties typically associated with the so-called ``small-world'' characteristic of the network \cite{Watts1998}, namely the shortest path lengths $L_{ji}$ between counties and the clustering coefficient of nodes $C_i$ (see definitions in \emph{Materials and Methods}). The shortest path length can be related to time scales for search or spreading processes, i.e. the time it takes to reach one location starting at another. The clustering coefficient quantifies the magnitude of the average flow between triplets of neighboring locations---a large value indicates that two neighbors of a location are likely to have large flows between them, too. Numerous systems are associated with high clustering while having small shortest paths (typically mean shortest paths scale logarithmically with systems size), which is referred to as the ``small-world'' property. This property typically facilitates the spread of epidemics \cite{Pastor-Satorras2015, Newman2002}. In contrast, lattices typically have shortest paths scaling polynomially with system size and high clustering and thus comparatively slower spreading speeds\cite{Barthelemy2011}. 

%The main structural changes we observe are as follows: During lockdown, both average shortest path length $L(T)$ and average clustering coefficient $C(T)$ increase substantially (see Fig.~\ref{fig:networks_measures}). Both properties are indicative of the small-world structure of a network \cite{Watts1998}. While $L(T)$ is a global property that measures the typical separation between nodes, $C(T)$ is a local property that measures the cliquishness of a typical neighborhood. The increase in both measures indicates that the network has simultaneously become more sparse on a global scale, and relatively more dense on a local level. The lockdown network is more lattice-like, with predominantly local connections and fewer connections between remote parts of the network, indicating a reduction of the system's ``small-world'' property.

We observe substantial changes in the structural properties of the mobility networks during lockdown, as illustrated by the shortest path trees for the weekly mobility networks $G_T$ (see Fig.~\ref{fig:networks_measures}A). In the pre-lockdown network $G_{10}$, long-distance connections enable a quick traversal of the network followed by few local steps. In the lockdown network $G_{13}$, the shortest paths are generally longer and include more local steps between neighboring counties. As a consequence of these structural changes, both the average shortest path length $L(T)$ and average clustering coefficient $C(T)$ increase substantially (see Fig.~\ref{fig:networks_measures}B). Moreover, we observe a striking difference in expected path length as a function of geographic distance, see Fig.~\ref{fig:networks_measures}C. In the pre-lockdown network, the expected shortest path length $L_{d}$ initially increases with geographical distance $d$, but eventually saturates to an almost constant niveau for $d\gtrsim550\,\mathrm{km}$, i.e.~is independent of geographic distance. This is a well known phenomenon of spatial small-world networks, where it has been shown that shortest path lengths typically scale as $L\propto r$ with Euclidian distance $r$ up to a critical distance $r_\mathrm{c}$, followed by an independence regime, $L\approx\mathrm{const.}$ for $r > r_\mathrm{c}$ \cite{Moukarzel1999}. A similar relation has been found in empirical human mobility networks such as air traffic networks \cite{Brockmann2013}. In such networks, geographic distance is an unreliable predictor for the effective arrival time because larger geographical distances can quickly be overcome by traveling along links connecting distant places. However, in the lockdown network, we observe a continued, almost linear dependence of the shortest path length on geographical distance, which is a typical property of lattices \cite{Barthelemy2011}. Because long-distance links are missing or weak, and travel predominantly occurs along short-distance connections, geographical distance dominates effective travel distance or travel duration.

We therefore conclude that the lockdown network is more lattice-like, with predominantly local connections and fewer connections between remote locations, reflecting a reduction of the system's ``small-world'' property. As indicated above, this has important implications for dynamical processes such as epidemic spreading, which we will discuss in the next section.

The unexpected scaling relation between path lengths and geographic distance during lockdown cannot merely be explained by the fact that the total flow is reduced in the lockdown network, neither is it due to thresholding effects. To demonstrate this, we use the rescaled pre-lockdown network $G_{10}^*(T=13)$ as a comparison. As we see in Fig.~\ref{fig:networks_measures}C, the reduced flow accounts for the changes at small distances, but it does not explain the different dependence of the shortest path length on geographical distance at high distances. This confirms that the observed effect is due to structural differences between the pre-lockdown and lockdown networks. In the Appendix section B, we present further evidence to support this conclusion by evaluating how several spreading time scales change over time in both the measured mobility networks as well as rescaled networks.

\section{Effect of lockdown on spreading processes}

\begin{figure*}[ht]
\centering
\includegraphics[width=1.0\linewidth]{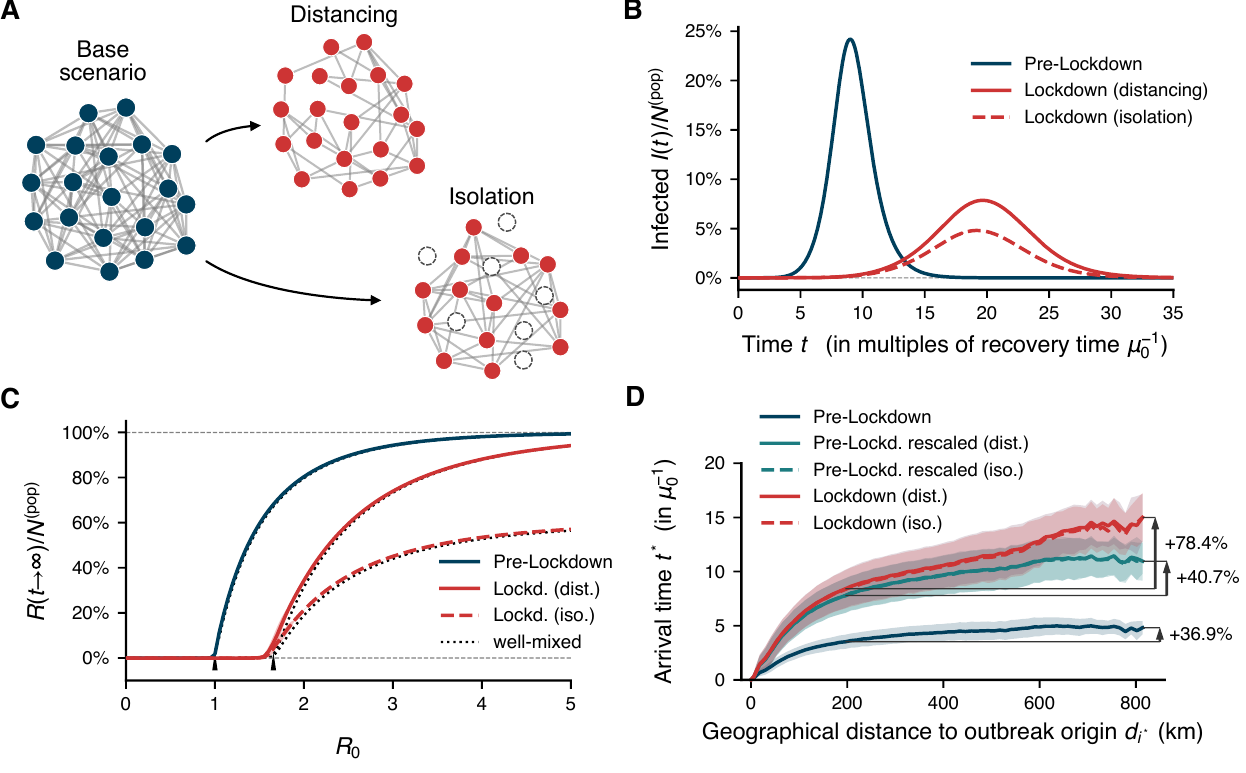}
% \caption{TEST}
\caption{\label{fig:epi_effects} Simulations of an SIR-epidemic on pre-lockdown and lockdown mobility networks. (A) We incorporate changes in total mobility in two scenarios in the model: In the ``distancing'' scenario, reduced mobility removes contacts between individuals, uniformly distributed over all individuals. In the ``isolation'' scenario, reduced mobility implies that an equivalent fraction of individuals isolate at home and are effectively removed from the system (see main text and SI for details). (B) In both model scenarios, the epidemic curve (infecteds over time) is flattened and its peak shifted to later times during lockdown. Note that we omit simulations on the rescaled network that yield similar results, indicating that the observed flattening effect is dominated by a decreasing basic reproduction number rather than structural changes. Results are shown for $\mathcal R_0=3$ and recovery rate $\mu=1/(8\mathrm{d})$, with a single random outbreak origin of $I_0=100$, averaged over 1000 simulations for each scenario. (C) The epidemic threshold is shifted to higher values of $\mathcal R_0$ during lockdown in both lockdown scenarios. Arrows indicate the well-mixed epidemic thresholds at $\mathcal R_0^*=1$ (pre-lockdown) and $\mathcal R_0^*=1.67$ (lockdown). The threshold is higher in the metapopulation model compared to the well-mixed description, an expected effect caused by the heterogeneity of the system \cite{Colizza2008}. (D) The average arrival times $t^*$ in counties as a function of geographic distance $d_{i^*}$ from the outbreak origin $i^*$. The arrival time $t^*$ is defined as the first time when infecteds pass the threshold of $0.1\%$ in a county. In the lockdown network, arrival times increase due to lower mobility. More importantly, however, we observe a similar scaling relationship as shown for the shortest path lengths: During lockdown, the arrival time shows a continued increase as a function of geographic distance from the outbreak origin, even in the long-distance regime. The changed scaling behavior cannot be explained by the lower total amount of trips (rescaled network).}
\end{figure*}

\subsection{SIR model with containment}

Finally, we address the question to what extent the lockdown-induced changes in mobility impact epidemic spreading processes mediated by the mobility network. We simulate an SIR epidemic metapopulation model \cite{Ray1997, Keeling2007}. In SIR models, individuals are assumed to be in either of three distinct states: susceptible (S), infected (I), or removed (R) from the transmission process. Contacts between susceptibles and infecteds may lead to the infection of the susceptible individual and infected individuals can spontaneously be removed from the transmission process by medical/non-medical interventions, death or immunization. In metapopulation models, infecteds based in one location can cause infections in other locations with a rate proportional to the daily flow between locations. Implicitly it is assumed that individuals travel back an forth and transport the infection between areas.

Note that in the following, we use epidemiological parameters similar to those of COVID-19 (see \emph{Materials and Methods}). However, we do not aim to replicate the actual spread of COVID-19 in Germany, but rather intend to demonstrate qualitative effects of the lockdown on epidemic spreading in general.

%The basic reproduction number $R_0$ quantifies the average number of secondary infections a single infected individual causes in an entirely susceptible population. The basic reproduction number is therefore the major control parameter that determines the total outbreak size as well as the growth speed of an epidemic. Furthermore, infected individuals spontaneously recover with recovery rate $\mu$. 

We implement a well-known commuter-dynamics SIR metapopulation model \cite{Tizzoni2014} with minor modifications. Specifically, the original model does not account for changes in the total amount of mobility (i.e.~total number of trips). The modified model accounts for the drastic reduction in total mobility, a substantial part of the changed mobility patterns due to containment measures.

%In addition, recent theoretical studies with SIR-like model have incorporated quarantine mechanisms and underlined their impact on epidemic spreading \cite{Maier2020, Arenas2020}. 

To include changes in the total amount of mobility in the model, we assume that a reduction in mobility reduces the rate with which contacts between infecteds and susceptibles cause infections. We implement this in two variants, to capture different methodological approaches: In the ``distancing'' scenario, mobility reduction leads to a proportional reduction in the average number of contacts. The ``isolation'' scenario instead implies that the equivalent percentage of the population isolates at home while the remaining individuals do not change their behavior (see Fig.~\ref{fig:epi_effects}A for an illustration, and \emph{Materials and Methods} and Appendix for details on the SIR model). Note that while many other non-pharmaceutical interventions may mitigate the spread of an infectious disease, we purely aim to discuss the effect of reduced and restructured mobility here.

\subsection{Mobility reduction flattens the curve}

An analysis of the SIR model indicates that lockdown measures have a distinct impact on epidemic spreading, see Fig.~\ref{fig:epi_effects}. Most prominently, a reduction of mobility reduces the overall incidence of the epidemic and delays its spread, shifting the peak to later times: The lockdown measures ``flatten the curve'' of the epidemic (Fig.~\ref{fig:epi_effects}B). This applies to both lockdown scenarios implemented here, where the stricter ``isolation'' scenario shows a lower overall incidence. The rescaled pre-lockdown network shows an almost identical incidence curve to the lockdown network (not shown here), which indicates that the ``flattening'' is mostly caused by the reduction in overall traffic.

In addition, lockdown measures increase the epidemic threshold $\mathcal R^*_0$ of the disease, that is the minimal force of infection required to infect a substantial amount of the population, see Fig.~\ref{fig:epi_effects}C. We compare the simulations to results of the canonical well-mixed model (see \emph{Materials and Methods}). We find little difference between the results of the metapopulation simulation and the well-mixed model, which suggests that stochasticity plays only as small role in the metapopulation system.

\subsection{SIR model replicates geographic dependence of arrival times}

An important observation is that the spread of the epidemic shows a similar functional dependence on geographic distances as the the shortest paths. This implies that the observed structural changes have considerable practical implications. To clarify this point we measured the arrival times of the epidemic in counties, see Fig.~\ref{fig:epi_effects}D. During lockdown, the epidemic takes longer to spread spatially, which is caused by the reduced contact numbers due to reduced mobility. More importantly, a stronger and continued increase of the arrival time with geographic distance from the outbreak origin during lockdown is observed: The farther away a county is from the outbreak origin, the longer it will take for the county to be affected by the epidemic. In contrast, with pre-lockdown mobility, the arrival times exhibit only a slow increase with geographic distance. Furthermore, the rescaled network does not replicate the changed scaling relation of arrival times during lockdown, which demonstrates that it is not caused by a reduction in the total amount of trips.

The dependence of arrival times on geographic distance in Fig.~\ref{fig:epi_effects}D matches the corresponding relationships for the shortest path lengths depicted in Fig.~\ref{fig:networks_measures}C. Therefore, structural changes---i.e. a reduced connectivity across long distances---have direct consequences for the dynamics of an epidemic, mitigating the spatial spatial spread over long distances.

\section{Discussion}
\label{sec:discussion}

In this study, we report and analyze various lockdown induced changes in mobility in Germany during the initial phase of the COVID-19 pandemic. We found a considerable reduction of mobility during the pandemic, similar to what was previously reported for other countries that passed and implemented comparable policies \cite{Klein2020, Lee2020, Pepe2020, Gao2020}. The reduction in mobility can be divided into a swift decrease, early in the lockdown phase, followed by a slow recovery. The initial rebound occured in late March although official policies remained unchanged. This could be indicative of individuals taking up non-essential trips again in spite of lockdown policies. We think that further research is necessary to illuminate what part of the mobility reduction was a direct consequence of policies, and which part was caused by voluntary behavioral changes within these official regulations.

We found evidence for profound structural changes in the mobility network. These changes are primarily caused by a reduction of long-distance mobility, resulting in a more clustered and local network, and hence a more lattice-like system. Most importantly, we found that path lengths continually increase with geographic distance, which is a qualitative change compared to pre-lockdown mobility. These changes indicate a reduction of the small-world characteristic of the network. 
%However, we should note that an extension of these measures to the case of weighted, directed flow networks as we study here is not straight-forward, such that the small-world phenomenon is not well-defined for these networks \cite{Saramaki2007, Muldoon2016}. In addition, the high density of the network studied here reduces the impact of the small-world phenomenon, as still relatively many long-range connections exist during lockdown. Despite these caveats, we think that the small-world concept is the most reasonable conceptual framework in which to frame our findings and that the comparison is valid.

In the context of human mobility, the structural network changes can be interpreted in different ways. Fewer individuals travel along connections of growing distance. One possible reason for this is that the individual ``cost'' of traversing long-distance connections has increased, more so than that for short-distance links, for example due to legal restrictions on travel, missing transportation options (such as flights), or slower and reduced transportation overall. As a result, people might avoid such travel or break up their travel in smaller trip segments.

The practical consequences of our findings are highlighted in the epidemic simulations analysis. We found that reduced global mobility during lockdown likely slowed down the spatial spread of the disease. Regarding structural changes, we found that the arrival times in counties increase continuously with the distance to the outbreak origin during lockdown, matching results of the topological analyses of the shortest path lengths. This result emphasizes our argument that the changes in the mobility network shown in this study have direct and non-trivial consequences on dynamic processes such as epidemic spreading. Our findings also suggest that targeted mobility restrictions may be used to effectively mitigate the spread of epidemics. In particular, measures that reduce long-distance travel mitigate a diseases’ spread during the first phase of an outbreak while a reduction in general mobility may be associated with a flattened prevalence curve.

In conclusion, we hope that future research will further illuminate the complex effects of restrictive policies on human mobility. Deeper and more complex aspects of mobility changes may occur during lockdowns, ranging from topological properties of the mobility network to its relation to sociodemographic and epidemiological conditions of the affected regions. We hope that a clearer understanding of complex effects of mobility-restricting policies will enable policy-makers to use these tools more effectively and purposefully, and thus help to mitigate the ongoing COVID-19 pandemic and to better prepare us for future epidemics.

%\matmethods{
\section{Materials and Methods}

\subsection{Daily mobility change}

To investigate national mobility trends, we focus on the total number of trips $N(t)=\sum_{i,j=1}^m F_{ji}(t)$ on the date $t$. In order to judge whether mobility has changed during the pandemic, we compare the mobility during the pandemic $\mathcal{T}$ to a baseline period with ``normal'' mobility  $\mathcal{T}_0$. Different comparison time frames $\mathcal{T}_0$ can be chosen and no clear, optimal choice that exhaustively accounts for seasonal effects, holidays, and general changes in mobility patterns exists. Here, we use March 2019 as a comparison, which we assume to be structurally closest to the period of March 2020 where most interventions took place. 

For a given date $t$ within the time frame of the pandemic $\mathcal{T}$, we calculate the mobility change $\Delta n(t)$ by comparing the number of trips $N(t)$ to the expected number of trips $N_0(t)$ during the baseline mobility period $\mathcal{T}_0$ as 
%
%\begin{linenomath}
$$ \Delta n(t) = \left( \frac{N(t)}{N_0(t)} \right) - 1.$$
%\end{linenomath}
%
% NEW VERSION:
Because mobility differs strongly depending on the weekday, we calculate the expected number of trips $N_0(t)$ as the average number of trips on all those dates $\mathcal D_\tau$ in the base period $\mathcal{T}_0$ that have the same weekday $\tau$ as the date $t$,

%\begin{linenomath}
$$ N_0(t) = |\mathcal D_\tau|^{-1} \sum_{t'\in\mathcal D_\tau} N(t').$$
%\end{linenomath}

In order to analyze the mobility change for a single county, $\Delta n^{(i)}(t)$, we use the same procedure but only count the number of trips that originate in the county $i$, i.e.~${N^{(i)}(t)} = \sum_{j=1}^m F_{ji}(t)$
%
% OLD VERSION:
%Because weekly modulations are most pronounced in the data, we choose the baseline number of trips $N_0(t)$ as the average over all weekdays of our base period $\mathcal{T}_0$ (March 2019) that correspond to the weekday of day $t$,
%
%$$ N_0(t) = |\Omega(t)|^{-1} \sum_{\tau\in\Omega(t)} N(\tau).$$
%
%Here, $\Omega(t)$ is the set of all days in March 2019 that correspond to the same weekday of day $t$, i.e.~given a function $\omega(t)$ that returns the weekday of any given day $t$, $\Omega$ is defined as
%
%$$\Omega(t) = \{ \tau \in \mathcal T_0: \omega(t) = \omega(\tau) \}.$$
%In order to analyze the mobility change for single counties, we measure the relative change in the number of outgoing trips as $\Delta n^{(i)}(t) = {N^{(i)}(t)}\big/{N^{(i)}_0(t)} - 1,$ where ${N^{(i)}(t)} = \sum_j F_{ji}(t)$ and $N^{(i)}_0(t) = |\Omega(t)|^{-1} \sum_{\tau\in\Omega(t)} N^{(i)}(\tau)$.

\subsection{Mobility change for distances}
When we calculate the distance-dependent mobility change $\Delta n_D(t)$, we proceed similarly to the previous section, but we only consider trips whose distance falls into a certain distance range $D=\{d: d_\mathrm{min} < d \leq d_\mathrm{max}\}$. As a proxy for the distance of flows $F_{ji}(t)$, we use the geographical distance $d_{ji}$ between the centroids of counties $i$ and $j$.
% NEW VERSION:
The number of trips in the distance range $D$ is
%\begin{linenomath}
$$N_D(t)=\sum_{(i,j)\in\Phi_{D}} F_{ji}(t).$$
%\end{linenomath}
where $\Phi_D$ is the set of all pairs of counties $(i,j)$ whose distance falls into the range $D$,
%\begin{linenomath}
$$ \Phi_D = \left\{ (i,j): d_{ji} \in D \right\}.$$
%\end{linenomath}
% OLD VERSION:
%Then, the set of flows within the distance range $D$ is given as
%$$\Phi_D(t) = \left\{ F_{ij}(t): d_{ij} \in D \right\},$$
%and the total number of trips in this distance range as
%$$N_D(t)=\sum_{f\in\Phi_{D}(t)} f.$$
Using $N_D(t)$, we calculate $\Delta n_D(t)$ as outlined in the previous section.

\subsection{Calculation of weekly mobility networks}
We create weekly mobility networks $G_T$ from trips measured during calendar week $T$. Let $\mathcal D_T$ denote the set of days in calendar week $T$. The edge weights $w_{ji}(T)$ are then calculated as the average daily number of trips between counties during this week,
%\begin{linenomath}
$$w_{ji}(T) = |\mathcal D_T|^{-1} \sum_{t'\in \mathcal D_T}F_{ji}(t').$$
%\end{linenomath}
We omit edges whose average weight is below the threshold $w_{ji}(T)<5$ to ensure consistency and comparability with the daily data.

\subsection{Rescaled networks}
To investigate how the global reduction of mobility affects our observations in comparison to structural changes, we construct rescaled networks $G_{10}^*(T)$ by scaling the weights of the pre-lockdown network of calendar week 10 by the flow lost during week $T$, i.e.~we set
%\begin{linenomath}
$$ w_{ji}^*(T) = w_{ji}(T=10) \times \frac{\sum_{i,j=1}^m w_{ji}(T)}{\sum_{i,j=1}^m w_{ji}(T=10)}.$$
%\end{linenomath}
Subsequently, we apply the same thresholding procedure as was done in the original data to the resulting network and discard all links with $w_{ij}^*(T)<5$. We therefore obtain a network that is structurally similar to the pre-lockdown system of calendar week 10 but has the same total amount of trips as the corresponding system of calendar week $T$, which allows us to isolate the effects that come purely from a uniform, global mobility reduction and subsequent thresholding.

\subsection{Path lengths and clustering coefficient}
To measure path lengths in the network, we consider two counties to be ``close'' to each other when they are connected by a large flow value and define the distance of each link as the inverse weight along the edge $\ell_{ji}=1/w_{ji}$.
%==============================
% This is what I would write alternatively:
%==============================
%The weight $w_ij(T)$ counts the average number of trips from $i$ to $j$ per day in calendar week $T$. Therefore, $\tilde w_{ij} = w_{ij} / \delta t$ with $\delta t = 1\mathrm{d}$ corresponds to an average transition rate of mobile phones from $i$ to $j$ in calendar week $T$. Thus, the ``length'' of an edge reflects the average waiting time until a transmission from $i$ to $j$ takes place in units of $\delta t$.
Using this distance metric we calculate the shortest path length $L_{ji}$ between each pair of source node $i$ and target node $j$ using Dijkstra's algorithm \cite{dijkstra1959note}. We calculate the weekly average path length $L(T)=(m(m-1))^{-1}\sum_{i,j=1}^m L_{ji}(T)$ and the average weighted and directed clustering coefficient $C(T)=m^{-1}\sum_{i=1}^m C_{i}(T)$ over all nodes for the weekly networks $G_T$ (as defined in \cite{Fagiolo2007}). Because the above definition of distance is sensitive to changes in the total flow of the network, we discuss a variety of other distance scales in the SI, yielding similar results (see Appendix section C2 and 3). Additionally, we show that increasing the observation threshold $w^c$ does not substantially change the results, indicating that the original threshold of the data was chosen small enough to not have an impact on our conclusions.

\subsection{SIR metapopulation model}

We use a modified version of the model proposed in \cite{Tizzoni2014} where susceptible $S$, infected $I$, and recovered individuals $R$ are associated to be part of commuter compartments $X_{ji}$ (with $X\in\{S,I,R\}$) when they live in location $i$ and work in location $j$. The compartments are coupled by shared work and home locations, respectively, and commuter-compartment population sizes $N^{\mathrm{pop}}_{ji}=S_{ji}+I_{ji}+R_{ji}$ are assumed to be proportional to the edge-specific outflux ratio of location $i$ as $N^{\mathrm{pop}}_{ji}=N^{\mathrm{pop}}_i F_{ji}/\sum_k{F_{ki}}$.  Full details are given in the Appendix section \ref{app:SIR}.

As stated in the main text, we incorporate two different variations of lockdown mechanisms into the model, to account for different interpretations of the influence of mobility reduction on the average number of contacts. In the ``distancing'' scenario, we assume that a mobility reduction by a factor $\kappa_{i}$ in a location $i$ leads to a linear reduction in the transmission rate $\beta_{i}$ throughout the epidemic, i.e.~$\beta^{'}_{i}=\kappa_{i}\times\beta$. The assumption here is that the reduced mobility uniformly translates into reduced contacts between individuals. In the other, stricter scenario ``isolation'', we instead assume that the reduced mobility means that individuals stop their commuting and are effectively removed from the system. We implement this by assuming that initially, a fraction $1-\kappa_{i}$ is removed from the transmission process such that $S^{'}_{ji}(t=0)=\kappa_{i} S_{ji}(t=0)$ and $R^{'}_{ji}(t=0)=(1-\kappa_{i}) S_{ji}(t=0)$. Both scenarios lead to a reduction of the basic reproduction number that is proportional to a reduction of mobility. In section \ref{subsec:mobility-change-to-number-of-contacts} of the Appendix, we argue that such a linear relationship corresponds to an upper bound of transmissibility reduction induced by mobility reduction.

Recent meta-reviews estimate the basic reproduction number $\mathcal R_0$ for COVID-19 in the range of 2-3 and the infectious period as roughly 7 days \cite{Alimohamadi2020, Park2020, To2020}. Accordingly, we use $\mathcal R_0=3$ and a recovery rate of $\mu=1/(8\mathrm{d})$, close to values previously used for the analysis of the disease's spread in Germany \cite{Dehning2020}.

%\subsection*{Epidemic threshold}
%Effectively, the epidemic threshold is scaled by the factor by which mobility is lost: If $R^*_0=1$ normally, during lockdown the epidemic threshold is $R^{'*}_0=(1/\kappa) \times R^*_0=1.67\times R^*_0$, as mobility is reduced to a fraction of $\kappa = 60\%$ during the lockdown week 13. 

\subsection{Data availability}
The mobile phone dataset is deposited in the Open Science Framework (OSF) data collection (\url{https://osf.io/n53cz/}) in an anonymized form, which will enable readers to replicate the main results of this paper (see SI for a description of the anonymization process). All other datasets used are publicly available: The ACAPS dataset on government policies, population data and county-level geodata for Germany. Their sources are listed in the Appendix section \ref{app:mobility_data_availability}. The Python code used for the SIR simulation is available at \url{https://github.com/franksh/EpiCommute} and included in the OSF repository. 

% }
% END OF MATERIALS AND METHODS

%\showmatmethods{} % Display the Materials and Methods section

\begin{acknowledgements}
We would like to thank Luciano Franceschina, Ilya Boyandin, and Teralytics for help regarding the mobile phone data. We also thank Vedran Sekara, Manuel Garcia-Herranz, and Annika Hope Rose for helpful comments regarding the analyses. B.F.M.~is financially supported as an \emph{Add-on Fellow for Interdisciplinary Life Science} by the Joachim Herz Stiftung.
\end{acknowledgements}

% \showacknow{} % Display the acknowledgments section

% \nocite{*}
\bibliography{covidmobility}
% \bibliography{sorsamp}% Produces the bibliography via BibTeX.

%%%%%%%%%%%%% APPENDIX %%%%%%%%%%%%%%%%%

\appendix

\section{Mobility dataset}

\subsection{Mobility data availability}
\label{app:mobility_data_availability}
An anonymized version of the dataset publicly available in the Open Science Framework (\url{https://osf.io/n53cz/}). The anonymized dataset enables readers to replicate the main results of our work. The anonymization is described in section~\ref{sec:anonymization}.

\subsection{Data collection and basic description}

The mobility dataset we study here is gathered from mobile phone logs from Telef\'{o}nica, a mobile phone provider with around $43.6$ million customers in Germany in 2019 \cite{Telefonica}, and aggregated by the company Teralytics. The resulting dataset contains mobility flows in Germany, namely the number of trips across and within counties on a given date. We use data recorded in the period of January 1 2020 up to June 10 2020, as well as data from March 2019, which we use as a comparison baseline for mobility changes during the COVID-19 pandemic. This study is the first to use this mobility dataset.

Trips are recorded from cell tower logs in the following way: A trip is started whenever a device leaves its current cell tower area A. The device might then pass through one or multiple other cell towers, until it becomes stationary again in cell tower area B. "Stationary" means that no further movement is recorded for approximately 15 minutes. The start- and end-towers A and B can be the same, such that self-loops are also recorded. Note that determining the actual location of a device from cell tower logs can be error-prone, for example due to fluctuations in signal strength, uneven spatial signal coverage, or the considerable variability in the size of cell tower coverage areas \cite{Barbosa-Filho2017}.

All movements are then spatially aggregated on the level of the 401 counties in Germany (corresponding to NUTS 3), and temporally aggregated daily. The end result is the mobility matrix $((F_{ij})(t)$, which contains the number of movements between all pairs of counties on that given day.

There is a thresholding applied to the data in the pre-processing: Flows $F_{ij}$ with less than 5 trips on a given day are not included in the data, due to data privacy reasons. This has some effects on the data, which we discuss in the main text.

\subsection{Data processing}

Here we describe some further processing of the mobility dataset we undertook. In the main text, we consider the distance of trips in multiple analysis. We estimate the geographical distance $d_{ij}^\mathrm{geo}$ of a trip $i\rightarrow j$ as the distance between the centroids of the counties $i$ and $j$. We want to point out that this proxy of distance is quite coarse, and we expect a significant stochastic error especially on small length scales. In addition, the centroid distance likely leads to a systematic underestimation of the trip length between counties that are subsumed within other counties for trips to and from the surrounding county. However, we expect both errors to have a small influence on medium and long distances, which is where our analysis is most focused on.

Furthermore, we have excluded dates with federal holidays in the period of observation from the data. We do this because federal holidays show a clearly abnormal mobility, which is generally very low when compared to the same average weekday from the baseline. In fact, the comparison to an average weekday does not reflect whether the mobility is actually higher or lower "than usual" on this holiday. In consequence, we omitted holidays to avoid a distortion of the data.

Finally, we have excluded the small county of Brandenburg an der Havel from the spatial maps in Fig.~\ref{fig:mobility_reduction}C and depict it in neutral white on the map. We have identified it as an outlier in the data with abnormally high mobility, which could for example be caused by changes in data collection by the mobile phone provider in that area. In the maps, it is very visible and distorts the perception of the image, focusing the attention on what we consider to be a statistical anomaly, which is why excluded it in the maps.

\subsection{Anonymization details}
\label{sec:anonymization}

The mobility dataset described above is publicly available in an anonymized form, which still enables readers to reproduce our main results.

First, we anonymized the identity of the 401 counties by replacing its NUTS3 designation with an integer ID, which remains fixed throughout the dataset. This anonymization leaves the structure of the network and all derived properties unchanged. As additional information we provide the centroid distance between each pair of counties, and the category of each county as shown in Fig.~1B of the manuscript (city, border, other).

Second, to hinder de-anonymization of the identities of the counties, we apply a small multiplicative noise to the centroid distances between the counties. For each pair of counties with non-zero distance, we multiply the centroid distance by a random number drawn from a normal distribution with $\mu=1$ and $\sigma=0.02$. We expect that this noise adds a negligible statistical error to the results.

We expect that all our results are reproducible with the anonymized dataset, with the exception of the geographic depictions of mobility in Fig.~1C and Fig.~3A.

\section{Other datasets}

\subsection{Data on policy measures and interventions}
% \label{app:acaps}
For structured information about non-pharmaceutical interventions, we use data aggregated by ACAPS \cite{ACAPS}, which can be downloaded at: \url{https://www.acaps.org/covid19-government-measures-dataset}. The dataset contains worldwide governmental policies that are issued in response to the COVID-19 epidemic, including policies affecting mobility for Germany. To collect only relevant policies that likely affect mobility, we analyzed the listed policies qualitatively and filtered for measures from these categories: limitations of public gatherings, border checks, border closures, school closures, partial lockdowns, and public services closures. For each measure it is listed whether it is an introduction of new measure or the phase-out of an existing measure.

\subsection{Population data}
Census population data for Germany on a county level (as used in the SIR model) was downloaded from \url{https://www.destatis.de/DE/Themen/Laender-Regionen/Regionales/Gemeindeverzeichnis/Administrativ/04-kreise.html}.

\subsection{Geo-information} County-level shapefiles for Germany, used to create graphics containing maps and to calculate distances between the centroids of counties as a proxy for trip distances, were downloaded at \url{https://ec.europa.eu/eurostat/de/web/gisco/geodata/reference-data/administrative-units-statistical-units/nuts}.

\section{Small-world observables}

\subsection{Clustering}

We use a definition of local clustering for weighted and directed networks as given by Eq.~(10) of \cite{Fagiolo2007}. For undirected binary networks of size $m$, the local clustering coefficient $C_i$ of a node $i$ is defined as the probability that two neighbors of a focal node are connected to each other, as well. For undirected weighted networks, this concept is extended to include triangle intensity
%\begin{linenomath}
$$
    I_{ijk} = (\hat w_{ij} \hat w_{jk} \hat w_{ki})^{1/3}
$$
%\end{linenomath}
where $\hat{w}_{ij} = w_{ij}/\mathrm{max}_{k\ell}\{w_{k\ell}\}$ is the weight of edge $(j,i)$ normalized by the maximum weight of the network. This implies that every existing triangle is compared to the maximally possible triangle intensity for which each contributing edge weight is equal to the maximum edge weight, in which case $I_{ijk} = 1$. The weighted local clustering coefficient is therefore proportional to the unweighted local clustering coefficient, modulated by the focal node's average triangle intensity as
%\begin{linenomath}
$$
C^{\mathrm{weighted}}_i={I_i}C_i,
$$
%\end{linenomath}
see ref.~\cite{Saramaki2007}. For directed networks, the number of triangles that can possibly exist for a triplet of nodes increases: Every triplet of nodes has a maximum possible number of eight triangles that can be formed (two options per node pair). Yet, considering a focal node $i$, only those edges that point to $i$ from other nodes or point from $i$ to other nodes can possibly form triangles. The weighted clustering coefficient therefore measures the existence probability and average intensity of triangles that can be spanned given a focal node's in- and outgoing connections.

\subsection{Shortest path lengths and other measures of temporal distance}

In unweighted networks, the shortest path length $L_{ji}$ quantifies the minimum number of steps necessary to traverse from node $i$ to node $j$ along the edges of the network \cite{Watts1998}. While this quantity is often viewed as a metric of spatial distance, it can also be interpreted as a metric of temporal distance, namely the minimal duration it takes a walker to traverse from node $i$ to node $j$ along the edges of the network.

In weighted networks, it is not straight-forward to define a distance between two nodes based on the weight $w_{ji}$ between them. Usually, as larger weights are associated with smaller distances, researchers choose a distance definition of $\ell_{ji} = 1/w_{ji}$ (see e.g.~\cite{Newman2001,Muldoon2016}). In our case, the edge weight $w_{ji}$ represents a flow by counting the number of people on a particular day that traverse from $i$ to $j$. It can therefore be interpreted as an activation rate for traversal from $i$ to $j$ in units of $1/\mathrm{d}$. Further following this picture, the distance $\ell_{ji}=1/w_{ji}$ represents the average waiting time for a $i\rightarrow j$ traversal event to occur. Assuming maximally random traversal processes, the average shortest path length
%\begin{linenomath}
$$L\equiv[m(m-1)]^{-1}\sum_{i\neq j}L_{ji},$$
%\end{linenomath}
therefore represents the average minimal mean first passage time between any nodes $i$ and $j$ in an edge-centered random walk on a network with of $m$ nodes \cite{Masuda2017}. We can illustrate this point by additionally investigating how other network-wide time-scales of networks with weights $w_{ji}$ change. The propagator of an edge-centered random walk is given by the unnormalized graph Laplacian
%\begin{linenomath}
$$\mathcal{L}_{ji}=\delta_{ji}\sum_k w_{ki}-w_{ji}.$$
%\end{linenomath}
The eigenvalues of this operator are real and non-negative with an ordering of $0=\lambda_1<\lambda_2\leq...\leq \lambda_N$ for a network consisting of a single component. The so-called relaxation time $t_{\mathrm{rlx}}=1/\lambda_2$ quantifies the time scale with which an edge-centered random walk approaches its uniform equilibrium distribution \cite{Masuda2017}. Note that both the average shortest path length $L$ as well as the relaxation time $t_{\mathrm{rlx}}$ are sensitive to changes in the global flow.

In contrast, we can also investigate time  scales of discrete-time node-centered random walks, where the propagator is given by the transition matrix $p_{ji} = w_{ji} / \sum_k w_{ki}$. Note that for this process, time represents the number of traversals (i.e. steps) between nodes. The operator $p_{ji}$ is a stochastic matrix with largest eigenvalues $\pi_m=1$ and second-largest eigenvalue $\pi_{m-1}$. The so-called mixing time $t_{\mathrm{mix}}=1/(1-\pi_{m-1})$ quantifies the time-scale with which a node-centered random walk process approaches its non-uniform equilibrium \cite{Mohar1997}. Since entries of this matrix represent transition probabilities and are therefore normalized by a node's total outflow, the mixing time is not as sensitive to changes in the global flow as the relaxation time.

\begin{figure*}[ht]
\centering
\includegraphics[width=0.9\linewidth]{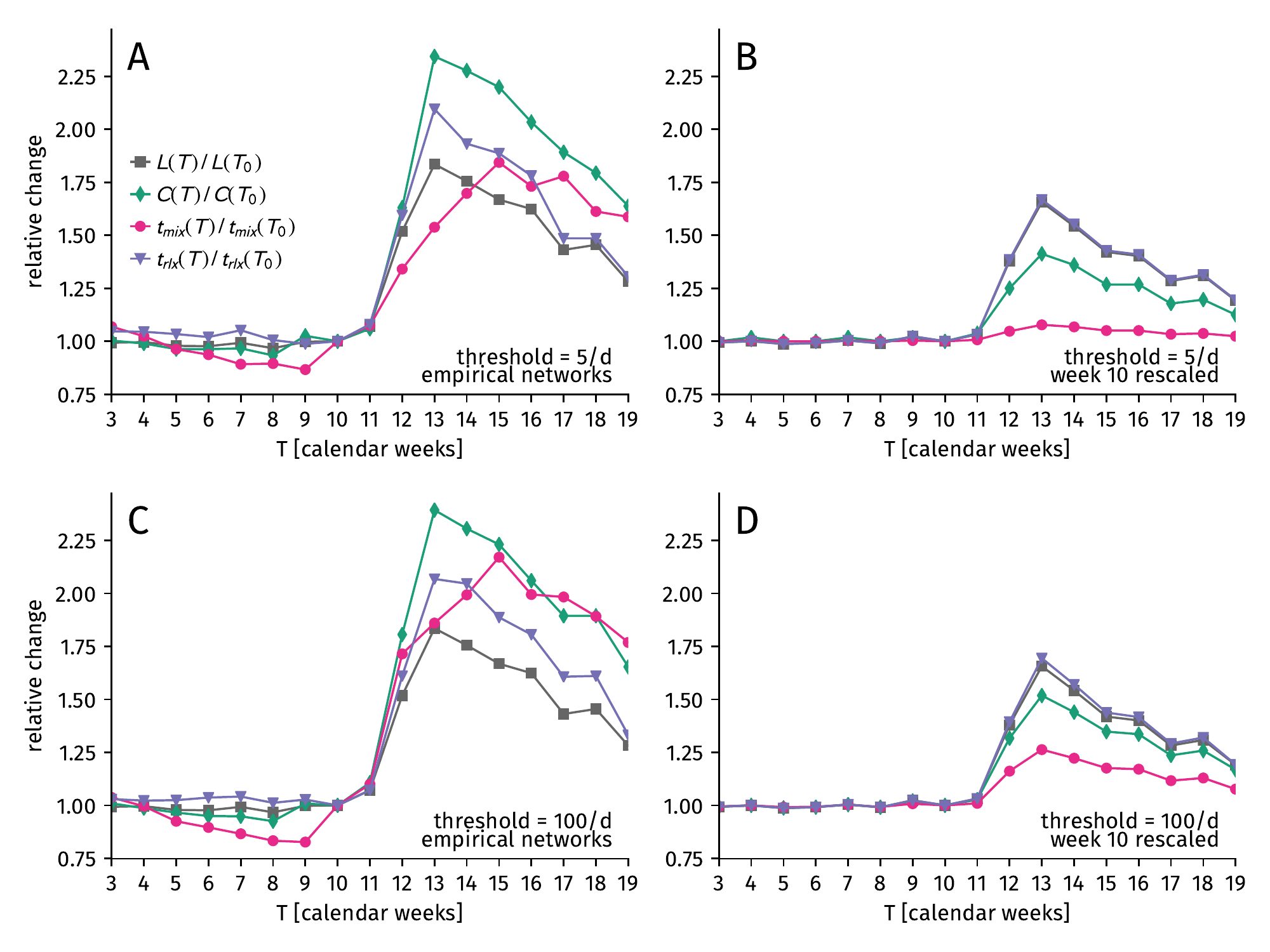}
\caption{\label{fig:small_world_other} Reduction of the small-world effect over time as illustrated by different global network observables. (A) The average clustering coefficient $C$, the average shortest path length $L$, the edge-centric random walk relaxation time $t_{\mathrm{rlx}}$, and the node-centric random walk mixing time $t_{\mathrm{mix}}$ all increase during lockdown. (B) Here, we show the same network observables for networks $G_{10}^*(T)$ that are topologically equal to the reference network $G_{10}$ of calendar week $T_0=10$ but where edge weights have been rescaled to sum to the total average flow observed in the corresponding calendar week $T$ (see Materials and Methods). We find that the relative increase of all network observables during lockdown cannot only be explained by a decreased total flow. In particular, the relative increase of the mixing time is heavily influenced by topological changes. (C) Network observables for networks that have been thresholded with $w_c=100$. (D) Observables for rescaled networks that have been thresholded with $w_c=100$.}
\end{figure*}

\subsection{Additional small-world analyses}

We want to investigate how the temporal observables discussed above (i) behave for the mobility networks averaged over calendar weeks and (ii) behave for networks $G^*_{10}(T)$ that are topologically equal to the network $G_{10}$ of calendar week $T_0=10$, but where edge weights have been rescaled to sum to the total average flow observed in the corresponding week $T$ (see the main manuscript, where this procedure was used to create the rescaled network $G^*_{10}(T=13)$ for calendar week $13$). The first exercise should show to what extent the shortest path length scales similarly to other network-wide spreading time scales in order to justify our choice of distance metric $\ell_{ji} = 1/w_{ji}$. Regarding the second point, we are interested in how much the reduction of the system's small-world property can be explained by a global reduction of trips.

We present our results in Fig.~\ref{fig:small_world_other}. As expected, the relaxation time $t_{\mathrm{rlx}}$ behaves similarly to the average shortest path length (compare Fig.~\ref{fig:small_world_other}A). Furthermore, the mixing time $t_{\mathrm{mix}}$ shows a relative increase of similar order during lockdown and subsequent transition to normal flow numbers, albeit of qualitatively different shape (compare Fig.~\ref{fig:small_world_other}A). As stated before, the mixing time is less sensitive to global flow modulations which indicates that the observed effect is indeed of topological nature. Investigating the observables on rescaled networks, we find that part of their increase during lockdown can be explained due to a global reduction in flow, but topological contributions cannot be neglected (compare Fig.~\ref{fig:small_world_other}B). Regarding the mixing time, almost all of its relative increase is explained by topological changes, as expected.

In order to investigate how strong our results are influenced by the thresholding procedure, we repeat the analyses for all networks after applying a higher threshold of $w_c=100$ (see Fig.~\ref{fig:small_world_other}C and Fig.~\ref{fig:small_world_other}D). We find that all relative observables but the mixing time remain virtually unchanged by an increased threshold, indicating that these results are rather stable. We do find a small increase for the mixing time, indicating that a small relative increase in this observable during lockdown might emerge due to thresholding effects.

\subsection{Additional visualizations of shortest path trees}

\begin{figure*}[ht]
\centering
\includegraphics[width=0.95\linewidth]{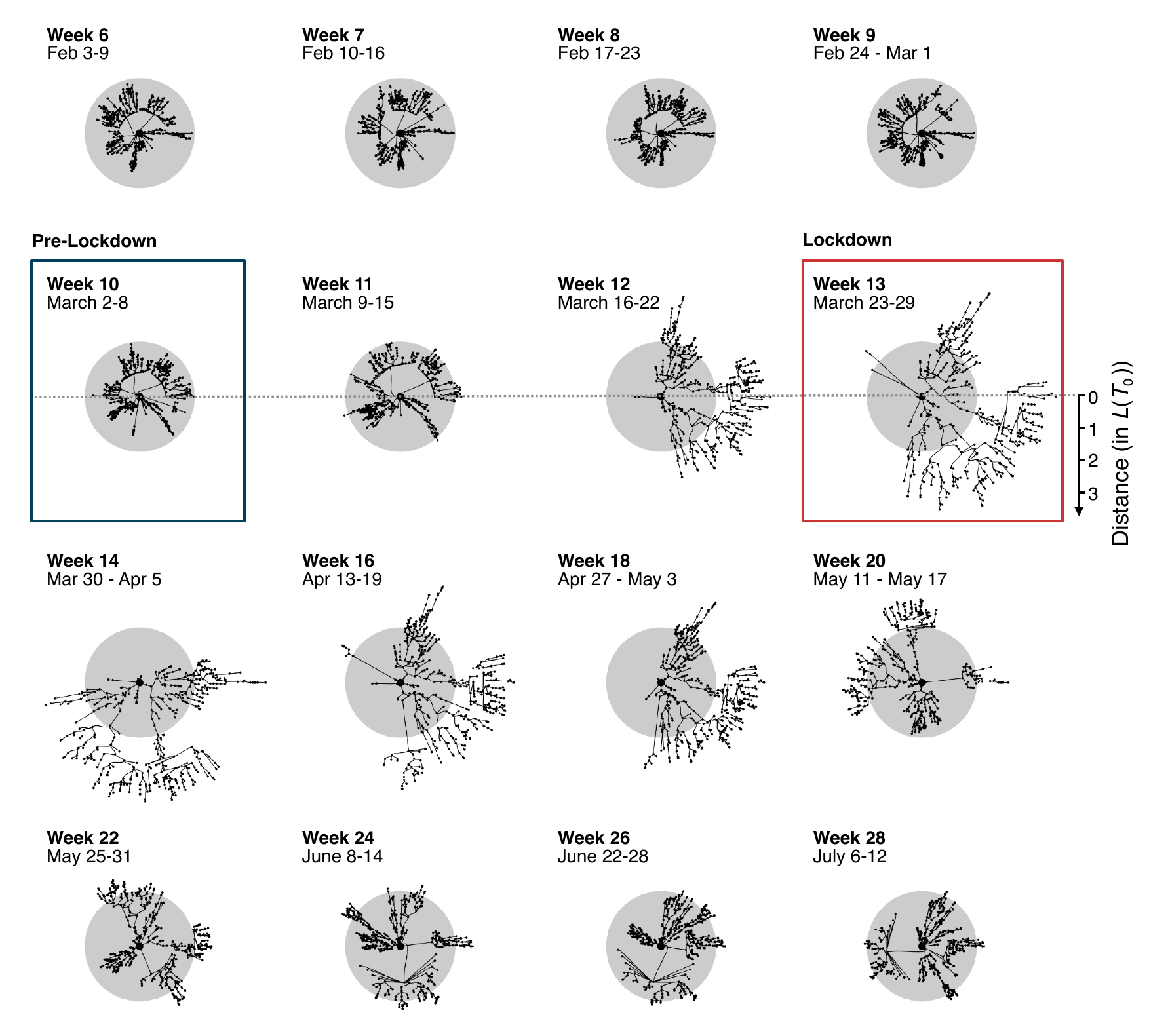}
\caption{\label{fig:berlin_spts} Shortest path trees for the weekly networks $G_T$, originating from Berlin. Structural changes can be observed starting in week $T=10$. Before the lockdown, shortest paths are relatively small and characterized by few jumps. During lockdown, shortest paths are longer and involve more. Grey circles show the maximum shortest path length $L_\mathrm{max}$ in week 10. The distance scale is in multiples of the average shortest path length $L(T_0)$ in the mobility network in week $T_0=10$.}
\end{figure*}

\begin{figure*}[ht]
\centering
\includegraphics[width=0.95\linewidth]{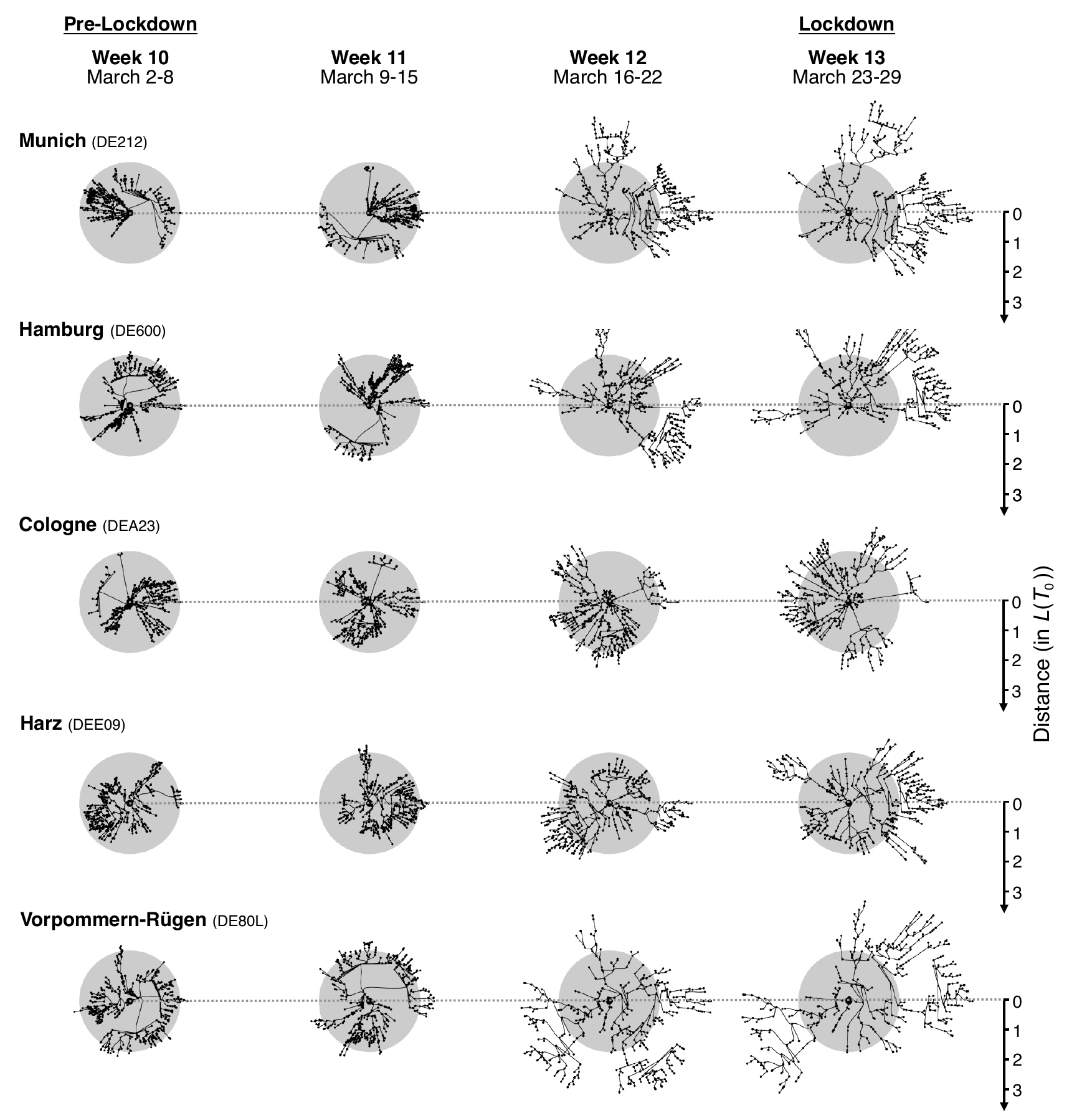}
\caption{\label{fig:other_sources_spts} Shortest path trees starting from different origins. We find the changes in the shortest path trees to be qualitatively similar irregardless of origin. Grey circles denote the maximum shortest path length $L_\mathrm{max}$ for the source Berlin in week $10$, for comparison. The distance scale is in multiples of the average shortest path length $L(T_0)$ in week $T_0=10$ for all sources.}
\end{figure*}

Here we provide additional visualizations of the shortest path trees of the weekly mobility networks $G_T$. The shortest path tree originating from Berlin (DE300), which is shown in Fig.~4A of the main text, is shown here for additional weeks, see Fig.~\ref{fig:berlin_spts}. In addition, we calculated the shortest path trees for other sources than Berlin, see Fig.~\ref{fig:other_sources_spts}.

In the pre-lockdown network originating in Berlin, paths are relatively short. The initial jumps in the paths are to only a few ``hub'' nodes, which act as gates to the other nodes. We find that these ``hub'' nodes are high-population counties with high traffic infrastructure (airports, train stations) that are geographically distant from Berlin, such as Munich, Frankfurt am Main, Hamburg or Cologne. These hubs then act as the entry gates to their regions. Using these hubs allows for quick travel over long distances, followed by short local paths.

We find that the greatest structural change occurs during weeks 10 and 13. During lockdown, paths become overall longer, and the hubs disappear, most likely due to the severe decline of train and airport and thus long-distance travel. Travel through the network occurs mostly along neighboring counties, which increases the lengths of paths. In the following weeks and months, the network slowly returns to its pre-lockdown state, although there are still structural differences persisting, both in that the paths remain longer and in that the hubs not yet regain the importance they had in the pre-lockdown network.

For the shortest path trees starting at other origins, we found their dynamics to be qualitatively similar to the ones starting in Berlin, see Fig.~\ref{fig:other_sources_spts}. All sources show an increase in their shortest path length during lockdown, a diminished importance of hubs and a more branching path tree.

\section{Details on the SIR Model}
\label{app:SIR}
We implement a metapopulation SIR-model \cite{Ray1997, Keeling2007} with commuter dynamics based on the model from \cite{Tizzoni2014}. An implementation of the model in Python is available in the Github repository \url{https://github.com/franksh/EpiCommute} and also stored in the OSF repository (\url{https://osf.io/n53cz/}). Our system is divided into $m$ counties with population $\Npop_i$, which we set to census population. Each counties' population is split into the compartments susceptibles $S_i$, infecteds $I_i$, and recovereds $R_i$, such that $\Npop_i=S_i + I_i + R_i$.

\subsection{Commuter mobility}

We assume that each individual has both a home location and a work location they commute to (which can also be the same county). The commuting patterns are measured by the mobility flow matrix $F_{ji}$. To compute the number of commuters for each pair of counties, we first construct the commute probability matrix $p_{ji}$ by normalizing the outgoing flows for each source district $i$,
%\begin{linenomath}
$$ p_{ji} = \frac{F_{ji}}{\sum_j F_{ji}},$$
%\end{linenomath}
such that 
%\begin{linenomath}
$$ \sum_j p_{ji}=1.$$
%\end{linenomath}
Further, we define
%\begin{linenomath}
$$\Npop_{ji}=p_{ji} \Npop_i$$
%\end{linenomath}
as a sub-population of location $i$ that lives in $i$ and commutes to $j$ for work, such that $\Npop_{i}=\sum_j \Npop_{ji}$ at all times.

\subsection{Infection dynamics}

We seed an initial infection by setting all individuals to be susceptible $S_{ji}=\Npop_{ji}$, except for an infection origin county $i^*$ (chosen uniform at random from the set of all locations), in which we distributed $I(t=0)=100$ infecteds among the compartments $I_{ji^*}$ proportional to their relative sizes $\Npop_{ji^*}/\Npop_{i^*}$ (such that $S_{ji^*}(t=0)=\Npop_{ji^*}-I_{ji^*}$).

There are two ways in which a susceptible of compartment $S_{ji}$ can get infected:
\begin{enumerate}
\item while being at home in compartment $i$, induced by all the infecteds present in $i$ (this includes commuters of locations $k$ to location $i$), and
\item while commuting at compartment $j$, by all the infecteds present in $j$ (this includes commuters of locations $k$ to location $j$).
\end{enumerate}
We assume that people spend half the time at home and half the time commuting. The dynamics for the compartment $S_{ji}$ are consequently given as 
\begin{eqnarray}
\frac{\mathrm{d}S_{ji}}{\mathrm{d}t} =&& -S_{ji}\left( \lambda_i^\mathrm{home} + \lambda_j^\mathrm{work} \right)\nonumber\\
=&& -S_{ji}\left( 
\frac{\beta}{2}\frac{\sum_k^m I_{ki}}{\sum_k^m \Npop_{ki}} + \frac{\beta}{2}\frac{\sum_k^m I_{jk}}{\sum_k^m \Npop_{jk}}
\right),\label{eq:commuter_S}
\end{eqnarray}
where the first term $\lambda_i^\mathrm{home}$ represents the force of infection while at home (where transmission can occur from all the infected at home in $i$), the second term $\lambda_j^\mathrm{work}$ is the force of infection while commuting (where transmission can occur from all the infected commuting to $j$). 
The dynamics for $I_{ji}$ follow analogously with an additional recovery term, i.e.
\begin{equation}
\frac{\mathrm{d}I_{ji}}{\mathrm{d}t} = - \mu I_{ji} + S_{ji} \left( \lambda_i^\mathrm{home} + \lambda_j^\mathrm{work} \right).\label{eq:commuter_I}
\end{equation}
Since the population size is constant, the third equation follows as $\mathrm dR_{ji}/\mathrm{d}t=\mu I_{ji}$.

\subsection{Stochastic simulation}
We use a stochastic binomial sampling algorithm to perform numerical simulations of the model. Given the above equations, the probability that an individual in compartment $S_{ji}$ becomes infected in the time interval $\left[t, t+\Delta t \right]$ due to the total force of transmission $\lambda_{ji} = \lambda_i^\mathrm{home} + \lambda_j^\mathrm{work}$ is

%\begin{linenomath}
$$
P(\Delta t; \lambda_{ji}) = 1 - \mathrm{e}^{-\lambda_{ji}\Delta t}.
$$
%\end{linenomath}

For this to be valid we have to choose $\Delta t$ small enough such that the time-dependent $\lambda_{ij}$ can be assumed as constant during $\Delta t$. 

Then, the number of individuals in compartment $S_{ji}$ that become infected during  $\left[t, t+\Delta t \right]$ is drawn from a binomial distribution with the probability $P(\Delta t; \lambda_{ji}),$

$$
\Delta \left( S_{ji}\rightarrow I_{ji}\right) \sim \mathrm{Binom}(S_{ji}(t), P(\Delta t; \lambda_{ji})).
$$

Similarly, the amount of infected in $I_{ji}$ that recover during this time is given by

$$
\Delta \left( I_{ji}\rightarrow R_{ji}\right) \sim \mathrm{Binom}(I_{ji}(t), P(\Delta t; \mu)).
$$

\subsection{Extension to include changes in mobility}
\label{subsec:SIR_extension}

The base model does not account for changes in mobility patterns as observed in reality, namely that the total number of trips decreases substantially (note that only relative changes in the flow are considered due to the use of the commute probability matrix instead of the flow matrix). It is important to consider the trip reduction $\kappa = N(T_L)/N(T_0)$ in the model because one can expect that a reduction in trips directly influences the average number of close-proximity contacts $k$, which in turn affects the basic reproduction number
%\begin{linenomath}
$$
    \mathcal R_0 = \tilde \beta k /\mu,
$$
%\end{linenomath}
where $\mu$ is the recovery rate per infected and $\tilde \beta$ is the infection rate per contact between a single infected and a single susceptible (note that in our description above, $\beta = \tilde \beta k$). 

However, finding the precise relation between the trip reduction $\kappa$ and the number of close-proximity contacts $k$ is an open problem. In section \ref{subsec:mobility-change-to-number-of-contacts}, we provide arguments that a relative change in the total number of trips $\kappa$ results in at least a corresponding linear decrease in the basic reproduction number and at most a quadratic decrease. In the following, we decide to implement lockdown scenarios with linear scaling, that can therefore be seen as upper bounds or ``pessimistic'' interpretations of the influence that mobility reduction has on mitigation.

We model a decrease of the total number of trips as two distinct scenarios, both of which are based on the assumption that each individual of a population of size $\Npop_{ji}$ contributes a constant average number of trips $f_{ji}$ to the total trip count, such that $F_{ji} = \Npop_{ji} f_{ji}$. 

In the first scenario (``isolation'') we assume that a fraction $\kappa_i$ of individuals at location $i$ isolates entirely, where $\kappa_i$ is the change in the total amount of in- and outgoing trips from location $i$,
\begin{equation}\label{eq:kappa_i_definition}
    \kappa_i = \frac{\sum_k ( F_{ki}(T_L) + F_{ik}(T_L) ) }{\sum_k ( F_{ki}(T_0) + F_{ik}(T_0) )}.
\end{equation}
The total number of trips is then determined only from individuals who do not isolate  $F_{ji}(T_L)=[\kappa_{i} \Npop_{ji}(T_L)] f_{ji}$. In contrast, a fraction $1-\kappa_{i}$ of $\Npop_{ji}$ contributes individual trip counts of $f_{ji}=0$ in this picture.
As these individuals are considered to be in isolation, they are effectively removed from the transmission process, which we model by assigning them to the $R_i$ compartment initially using the adjusted initial conditions of 
%\begin{linenomath}
\begin{equation*}
\begin{split}
S_{ji}'(t=0) &= \kappa_{i}\times S_{ji}(t=0)\\
I_{ji}'(t=0) &= I_{ji}(t=0)\\
R_{ji}'(t=0) &= (1-\kappa_{i})\times S_{ji}(t=0).
\end{split}
\end{equation*}
%\end{linenomath}
Note that from Eq.~(\ref{eq:commuter_S}) it follows that the initial spreading rate (the initial effective reproduction number, respectively) therefore scales linearly with $\kappa_{i}$ (as motivated in the following section).

In the second scenario (``distancing''), we instead implement a mechanism where a reduction in mobility leads to a reduction in the transmission rate $\beta$. As the reduction in mobility $\kappa_i$ is location-specific (as defined in Eq.~\ref{eq:kappa_i_definition}), this results in a location-specific transmission rate $\beta_i=\kappa_i\beta$, which we add in Eqs.~(\ref{eq:commuter_S})~and~(\ref{eq:commuter_I}) by setting
\begin{align*}
    \lambda_i^\mathrm{home} &= \frac{\beta_i}{2}\frac{\sum_k^m I_{ki}}{\sum_k^m \Npop_{ki}} \\
    \lambda_j^\mathrm{work} &= \frac{\beta_j}{2}\frac{\sum_k^m I_{jk}}{\sum_k^m \Npop_{jk}}.
\end{align*}

To motivate this definition, we assume that a reduction in mobility in the commuter-compartment $i\rightarrow j$ translates into a homogeneous reduction of the individual trip count $f_{ji}$ (such that $F_{ji}(T_L) = \Npop_{ji}(T_L) [\kappa_i f_{ji}]$). If we further assume that an individual has, on average, $c_{ji}$ close-proximity contacts per trip and zero contacts if they do not move, the average close-proximity contact number per individual scales as $k_{ji} \propto c_{ji} f_{ji}$. A reduction in the individual trip count directly translates into a proportional reduction in contacts. Note that this is a conservative assumption, though---realistically, the number ${c_{ji}}$ of close-proximity contacts per trip would not remain constant: When $f_{ji}$ decreases, we expect $c_{ji}$ to decrease, too, because the density of people that can encounter each other during a trip is reduced. We therefore consider the linear relationship to be a ``pessimistic'' assumption. A more elaborate argumentation concerning this point is given in Appendix section \ref{subsec:mobility-change-to-number-of-contacts}.

Note that both scenarios yield the same effective reproduction number $\mathcal R_{\mathrm{eff}}=\mathcal R_0\kappa$ at $t=0$. For a well-mixed system, this implies that $I^{(\mathrm{iso})}(t)=\kappa I^{(\mathrm{dist})}(t)$ and $R^{(\mathrm{iso})}(t\rightarrow\infty)=\kappa R^{(\mathrm{dist})}(t\rightarrow\infty)$, since both have the same reproduction number but differ in effective population size by a factor of $\kappa$.

%change: If the mobility drops to half, the number of contacts is reduced by half. To this end, we calculate the edge-wise change in mobility
%$$\kappa_{ij}(T) = F_{ij}(T)/F_{ij}(T_0)$$
%and include the mobility change as a linear scaling factor in the dynamics:

%$$ \frac{\mathrm{d}S_{ij}}{\mathrm{dt}} = -\beta\times \kappa_{ij}(T)\times \ S_{ij}\left( 
%\frac{1}{2}\frac{\sum_k I_{ik}}{\sum_k N_{ik}} + \frac{1}{2}\frac{\sum_k I_{kj}}{\sum_k N_{kj}}
%\right).$$

\section{Relationship between mobility change and average number of contacts}
\label{subsec:mobility-change-to-number-of-contacts}
In the following, we provide arguments that the average number of close-proximity contacts ${k_{i}}$ for week $T$ (and therefore the basic reproduction number) decreases at least linearly with the total number of observed trips $F_{i}(T)$ (and thus with the trip reduction $\kappa_{i}(T)$), which we use as an assumption for the ``distancing'' scenario in Section 4\ref{subsec:SIR_extension}.
Note that in the following, observables refer to a single sub-population $i$ in calendar week $T$ in the sense of the SIR model discussed in Section 4, yet, to enhance readability, we abstain from using subscripts and use time-dependence only when necessary, which implies that e.g.~$F\equiv F_{i}(T)$, $k\equiv k_i$ etc.

%Consider a single population of $\Npop$ individuals where $F(T)$ quantifies the total number of trips recorded within a period of time of length $\Theta$ (e.g.~$\Theta=24\,\mathrm{h}$ for a single day).
We are interested in the relative change of the average number of close-proximity contacts $x = k_L/k_0$ in dependence of the relative change in total number of recorded trips $\kappa = F(T_L)/F(T_0)$ when lockdown measures are introduced. To this end, we develop a simple model.

First, we assume that individuals are in either of two states, active (A) or inactive (X), and that they transition between being active and being inactive with activity rate $\alpha$ and inactivity rate $\xi$. This implies that on average, an individual stays active for time $\tau_a=1/\xi$ (remains inactive for time $\tau_x=1/\alpha$, respectively) and that in equilibrium, the expected number of active individuals is given as
%\begin{linenomath}
$$
    A^* = \frac{\alpha}{\alpha+\xi} \Npop.
$$
%\end{linenomath}
We further assume that individuals can only be in contact with each other if they are active. Inactive individuals are considered to always be isolated.

Assuming that only contacts between active individuals exist and that the process is in equilibrium, the average total number of contacts in the system can be approximated as 
%\begin{linenomath}
$$
    E =\frac{p}{2} A^*(A^*-1)\approx\frac{p}{2}(A^*)^2.
$$
%\end{linenomath}
where $p$ is the probability with which each single possible pair of contacts exists. Given the handshaking theorem $k=2E/N$, we find that the average number of contacts per person scales as
%\begin{linenomath}
$$
    x = \frac{k_L}{k_0} \approx \left(\frac{A^*_L}{A^*_0}\right)^2.
$$
%\end{linenomath}
Note that isolation of an individual does not imply inactivity: A person can be in an active state but still have no contacts (if $p$ is sufficiently small).

Regarding the number of observed trips, we assume that it is proportional to the temporal integral of every individual's activity rate
%\begin{linenomath}
$$F \propto \Npop \alpha,$$
%\end{linenomath}
which is motivated as follows: An individual becomes active when they move, at which time they may come in contact with other individuals. When the individual becomes inactive, they likely stopped their movement (they may have stopped their movement before), which concludes the trip that is consequently counted as an increase in the number of trips $F$. Hence, movements are only recorded as soon as an individual becomes active and two consecutive trips have to be considered as being intersected by an inactive period. The total number of events at which an individual becomes active is therefore proportional to the temporal integral of the activity rate, $T\alpha$.
From this assumption we conclude that
%\begin{linenomath}
$$
\kappa = \frac{F(T_L)}{F(T_0)} = \frac{\alpha_L}{\alpha_0}.
$$
%\end{linenomath}
Furthermore, we assume that the average time an individual is active is independent of the implementation of lockdown measures, which is motivated by the fact that necessary trips such as grocery shopping and commuting to and from the work place do not decrease in duration; rather, we expect the total number of such trips to decrease. This implies that $\tau_a=1/\xi=\mathrm{const.}$ while the duration of inactivity $\tau_x=1/\alpha$ can change, i.e.~increase during lockdown. This implies that the change of the average number of contacts is given as
%\begin{linenomath}
$$
    x = \left(\frac{\alpha_L}{\alpha_0}\right)^2\left(\frac{\xi+\alpha_0}{\xi+\alpha_L}\right)^2.
$$
%\end{linenomath}
Note that $A^*<\Npop/2$ if $\alpha < \xi$, i.e. if the average time of being active is lower than the average time of being inactive, less than half of the population is in an active state. Mirroring real systems, we can assume that this inequality holds when averaged over a single day, because (i) most individuals are isolated at night and (ii) many individuals are not in contact with other people for the majority of the day \cite{cattuto2010dynamics, Stopczynski2014}.
%Most likely, it holds true for lockdown such that $\alpha_L < \xi$. 
Thus, we normalize activity rates by the constant inactivity rate as $y_0=\alpha_0/\xi$ and $y_L=\alpha_L/\xi$ such that $y_{0},y_{L}\in[0,1]$ and $y_L=y_0\kappa$. In total, we find
%\begin{linenomath}
$$
    x = \kappa^2 \left(\frac{1+y_0}{1+\kappa y_0}\right)^2.
$$
%\end{linenomath}
From this relationship, we see that a linear reduction $x_u=\kappa$ is an upper bound of the true reduction and a quadratic reduction $x_l=\kappa^2$ is a lower bound of the true reduction, i.e.

%\begin{linenomath}
$$
    \left(\frac{F(T_L)}{F(T_0)}\right)^2 \leq \frac {\mathcal R_{0,L}}{\mathcal R_{0,0}} \leq \frac{F(T_L)}{F(T_0)},
$$
%\end{linenomath}
where we used ${\mathcal R_{0}}\propto k$.

The proofs are straightforward. We begin with assumption
%\begin{linenomath}
\begin{align}
    \kappa \geq \kappa^2\left(\frac{1+y_0}{1+\kappa y_0}\right)^2.
\end{align}
%\end{linenomath}
This inequality is met for $\kappa=0$. For $\kappa>0$, we find
%\begin{linenomath}
\begin{align}
    1 &\geq \kappa\left(\frac{1+y_0}{1+\kappa y_0}\right)^2 \\
    1 + 2\kappa y_0 + \kappa^2 y_0^2 &\geq \kappa + 2 \kappa y_L + \kappa y_0^2 \\
    1-\kappa &\geq \kappa (1-\kappa) y_0^2
\end{align}
%\end{linenomath}
This inequality is met for $\kappa=1$. For $\kappa <1$ we find
%\begin{linenomath}
\begin{align}
   1 &\geq \kappa y_0^2.
\end{align}
%\end{linenomath}
which is met for $0<\kappa<1$ and $y_0\in[0,1]$, therefore the initial assumption holds true, q.e.d.

The inequality
%\begin{linenomath}
\begin{align}
    \kappa^2 \leq \kappa^2\left(\frac{1+y_0}{1+\kappa y_0}\right)^2
\end{align}
%\end{linenomath}
is met for $\kappa = 0$. For $\kappa > 0$ we find
%\begin{linenomath}
\begin{align}
    1+2\kappa y_0 +\kappa^2 y_0^2 &\leq 1+2y_0+y_0^2. \\
    2\kappa y_0 +\kappa^2 y_0^2 &\leq 2y_0+y_0^2
\end{align}
%\end{linenomath}
This inequality is met if both $\kappa y_0 \leq y_0$ and $\kappa^2 y_0^2 \leq y_0^2$. These are both met for $\kappa\leq1$ and $0\leq y_0\leq 1$, therefore the assumption holds true, q.e.d.
%%% Add this line AFTER all your figures and tables

%%%%%%%%%%%%% END APPENDIX %%%%%%%%%%%%%

\end{document}